\documentclass[12pt,a4paper]{article}
\pdfoutput=1
\usepackage[utf8]{inputenc}
\usepackage[T1]{fontenc}
\usepackage[margin=1.9cm]{geometry}
\usepackage[pdfstartview={FitH},bookmarks=false,linktoc=page,colorlinks=true,linkcolor=blue,citecolor=blue,urlcolor=blue]{hyperref}
\usepackage[numbered]{bookmark}
\usepackage{mathtools}
\usepackage{amssymb}
\usepackage{graphicx}
\usepackage[font=small,labelfont=bf]{caption}
\usepackage{authblk}
\usepackage{cite}
\usepackage{dsfont}
\usepackage{xcolor}
\usepackage[titles]{tocloft}
\usepackage{float}
\usepackage{subcaption}
\usepackage{color}
\usepackage{tikz}
\usepackage{ifthen}
\usepackage[warn]{textcomp}
\numberwithin{equation}{section}
\allowdisplaybreaks
\usepackage{multirow}
\usepackage{bbold}

\usetikzlibrary{patterns}

\usepackage{comment}

\usepackage{pgfplots}
\pgfplotsset{compat=1.9}

\usetikzlibrary{arrows.meta, patterns, intersections, backgrounds}
\usepgfplotslibrary{fillbetween}

\tikzset{ntp/.style={circle, thin, minimum size=2mm, inner sep=0,
fill=white,#1}}

\usepackage{caption}
\usepackage{subcaption}

\usepackage[normalem]{ulem}
\usepackage{soul}

\begin{document}
\title{\vspace{2cm}\textbf{ Complexity of Quantum Charged Particle in External Magnetic Field }\vspace{1cm}}

\author[]{M. Radomirov}

\affil[]{\textit{Department of Physics, Sofia University,}\authorcr\textit{5 J. Bourchier Blvd., 1164 Sofia, Bulgaria}

\vspace{20pt}\texttt{radomirov@phys.uni-sofia.bg }\vspace{0.1cm}}
\date{}

\maketitle

\begin{abstract}
In this paper, we investigate the circuit complexity of a quantum charged particle in an external magnetic field. Utilizing the Nielsen approach, we determine the complexity of thermofield double states as functions of time, temperature, and cyclotron frequency. We analyze both the complexity and the amplitude of its oscillations across various parameter values, and reveal that these results cannot be derived as a limit of the harmonic oscillator case. Finally, we calculate the rate of complexity and show that it obeys the Lloyd bound.
\end{abstract}

\tableofcontents

\section{Introduction}

In the past several years, the interplay between information theory and fundamental physics has attracted growing attention, particularly within the framework of AdS/CFT correspondence. Following the conjecture by Ryu and Takayanagi \cite{Ryu:2006bv, Ryu:2006ef}, which connects the entanglement entropy of a conformal field theory with the geometry of the corresponding anti-de Sitter spacetime, the relationship between gravity and entanglement has propelled significant advancements in the field.

However, recent studies indicate that entanglement alone is insufficient to capture all aspects of the bulk theory \cite{Susskind:2014moa, Susskind:2014rva}. A notable example is the inadequacy of entanglement entropy in describing the evolution of an eternal two-sided AdS black hole \cite{Maldacena:2001kr}. Specifically, in the context of a thermofield double state (TFD) dual to such a black hole, the entanglement entropy of the TFD reaches equilibrium \cite{Hartman:2013qma}, while the volume of the black hole's interior continues to grow well beyond this thermalization time. To address this matter, Susskind and collaborators \cite{Susskind:2014rva, Brown:2015bva, Couch:2016exn} introduced the concept of complexity, which can explore the growth of the black hole beyond the thermalization time of entanglement entropy. Their original proposal posits that for an eternal black hole, the complexity is proportional to the spatial volume of the Einstein-Rosen bridge connecting the two boundaries.

Defining complexity within the holographic framework involves intricacies that have resulted in various proposals, such as complexity equals action (CA) \cite{Brown:2015bva}, complexity equals volume (CV) \cite{Susskind:2014rva}, complexity equals spacetime volume (CV2.0) \cite{Couch:2016exn}, and complexity equals anything \cite{Belin:2021bga, Belin:2022xmt, Jorstad:2023kmq, Myers:2024vve}. These conjectures aim to understand the quantum computational complexity of states in a boundary theory and their correspondence to gravitational descriptions in the bulk. The inherent subtleties are inevitable due to similar ambiguities arising in complexity theory\footnote{For more information, see \cite{Belin:2021bga}.}.

The concept of complexity originally comes from computer science \cite{arora2009computational, moore2011nature}, where it quantifies the number of operations needed to complete a specific task using a given set of simple operations. Expanding on this idea, various forms of complexity have been developed, such as quantum complexity \cite{Nielsen:2005mkt, nielsen2010quantum, Dowling:2006tnk, Nielsen:2006cea}, time complexity \cite{cormen2001introduction}, and holographic complexity \cite{Susskind:2014rva, Brown:2015bva, Couch:2016exn}. Generalizing the notion of complexity to different systems requires careful consideration and assumptions about these systems. For example, in quantum computing, complexity is defined as the minimum number of simple unitary operations needed to transform one state into another.

Importantly, there is no single, universal definition of complexity. Rather, there exists a family of complexity measures that may be multiplicatively related under certain conditions. This idea aligns with Nielsen's concepts of the geometry of computations, or complexity geometry \cite{nielsen2010quantum, Dowling:2006tnk, Nielsen:2006cea}. In order to implement Nielsen's approach, various methods and techniques have been developed, such as the covariance matrix method \cite{Chapman:2018hou, Ghasemi:2021jiy, Doroudiani:2019llj, Khorasani:2023usq}, the Fubini-Study metric \cite{Chapman:2017rqy}, and others \cite{Bhattacharyya:2018bbv, Ali:2018fcz, Jefferson:2017sdb}.

Motivated by the significance of complexity in holography and other fields, we investigate the thermofield double (TFD) state of a quantum spinless charged particle moving in an external magnetic field. Specifically, we analyze the effect of the magnetic field on Nielsen complexity using the covariance matrix approach. This approach is versatile and can also be applied to study similar aspects in more general systems with various properties \cite{Bagchi:2001dx, Jafarov:2013cza, Fatyga:1990wx,Mandal:2012wp, Jing:2009lfc, BenGeloun:2009hkc, Heddar:2021sly,Boumali:2020fqd, Falek:2017amp, Nagiyev:2023fwk, Martinez-y-Romero:1995grr,Mannheim:2004qz, Masterov:2015ija, Dimov:2016vvl, Pramanik:2012bh,Bouguerne:2023wyl, Hamil:2022uwy, Ballesteros:2022bqx}.

The structure of this paper is as follows. In Section \ref{sec2}, we quantize the charged particle in an external magnetic field following the standard quantum mechanical recipe. In Section \ref{sec3}, we construct the thermofield double (TFD) states of the system and represent them in a suitable operator form, following closely \cite{chapman2019complexity, Avramov:2024kpt}. In Section \ref{sec4} we construct the thermal covariance matrix of the system. In Section \ref{sec5}, we compute the Nielsen complexity of the time-dependent TFD states and examine its properties in relation to temperature and the magnetic field. In Section \ref{sec6}, we calculate the complexity rate and show that it satisfies the Lloyd bound. Our findings are summarized in Section \ref{sec7}.

\section{Quantization of charged particle in an external magnetic field}\label{sec2}
In this section, we examine the Schrödinger equation for a charged particle in an external magnetic field. We find the analytic solutions and introduce two types of creation and annihilation operators to construct the Fock space. Later we will use the energy spectrum of the Hamiltonian to construct the TFD states.

\subsection{The wave function}
Let us consider a spinless charged particle in a homogeneous magnetic field \( B \) aligned with the \( z \)-axis. Restricting the motion to the \( xy \)-plane (\( p_z = 0 \)), the Hamiltonian takes the form:
\begin{align}
H=\frac{1}{2m} \big(\vec p -e\vec A\, \big)^2 = \frac{1}{2m} \big(p_x^2+p_y^2 \big) -\frac{eB}{2m}\big(xp_y -yp_x\big) +\frac{e^2B^2}{8m} \big(x^2+y^2\big),
\label{Hamiltonian-1}
\end{align}
where the vector potential \(\vec{A}\) is expressed in the symmetric gauge $\vec A=\frac{B}{2}(-y,x,0)$. Introducing the cyclotron frequency \(\omega = \dfrac{eB}{m}\) and utilizing the standard momentum \( p_i = -i \hbar \partial_i \), we obtain:
\begin{equation}\label{Hamiltonian1}
H=-\frac{\hbar^2}{2m}(\partial_x^2 +\partial_y^2) +\frac{i\hbar\omega}{2} \big( x\partial_y -y\partial_x \big) +\frac{m\omega^2}{8}  \big(x^2+y^2\big).
\end{equation}
It is useful to change to polar coordinates:
\begin{align}\label{PolCoord}
x=\lambda \rho\cos\varphi, \quad y=\lambda \rho\sin\varphi,
\end{align}
where \(\rho\) and \(\varphi\) are dimensionless, and \(\lambda\) is a length parameter. Hnce, the corresponding Laplacian and angular momentum become:
\begin{align}\label{DL}
&\Delta= \partial_x^2 +\partial_y^2= \frac{1}{\lambda^2} \bigg( \partial_\rho^2 +\frac{1}{\rho}\partial_\rho +\frac{1}{\rho^2}\partial_\varphi^2 \bigg),\quad L_z=-i\hbar \big(x\partial_y -y\partial_x \big) = -i\hbar\partial_\varphi.
\end{align}
The Schrödinger equation \( H \Psi(\rho, \phi) = E \Psi(\rho, \phi) \) in is:
\begin{equation}\label{SchrEq}
\bigg[-\frac{\hbar^2}{2m\lambda^2} \bigg( \partial_\rho^2 +\frac{1}{\rho}\partial_\rho +\frac{1}{\rho^2}\partial_\varphi^2 \bigg)  +\frac{i\hbar\omega}{2} \partial_\varphi + \frac{m\omega^2}{8} \lambda^2 \rho^2 \bigg] \Psi= E \Psi.
\end{equation}
The Hamiltonian is independent of \(\varphi\), so the angular momentum is conserved and associated with a magnetic quantum number \(\ell\). We use a standard ansatz to separate the variables:
\begin{equation}
\Psi(\rho,\varphi)=e^{i \ell\varphi}R(\rho), 
\end{equation}
which results in the following equation for the radial part $R(\rho)$:
\begin{equation}\label{Rro}
R''(\rho)+\frac{1}{\rho}R'(\rho) +\bigg( \frac{2m\lambda^2E}{\hbar^2}  +\frac{m\omega\lambda^2\ell}{\hbar} -\frac{\ell^2}{\rho^2} -\frac{m^2\omega^2}{4\hbar^2} \lambda^4\rho^2 \bigg) R(\rho)=0.
\end{equation}
Since $\lambda$ is just a scale factor in the coordinate transformation \eqref{PolCoord}, we have the freedom to choose it such as $\lambda=\sqrt{\dfrac{2\hbar}{m\omega}}$. Aditionaly we set $\rho^2 = r$, hence the radial Eq. \eqref{Rro} becomes 
\begin{equation}
rR''(r) +R'(r) +\bigg( \frac{E}{\hbar\omega} +\frac{\ell}{2} -\frac{\ell^2}{4r} -\frac{r}{4}\bigg) R(r)=0,
\end{equation}
which is a Laguerre type equation with a solution
\begin{equation} 
R_{n,\ell}(r) =r^{\frac{\ell}{2}} e^{-\frac{r}{2}} L_n^{(\ell)}(r).
\end{equation}
The relation between the principal quantum number $n$ and the energy is given by
\begin{equation}\label{E_n}
E_n= \hbar \omega \bigg(n+\frac{1}{2}\bigg).
\end{equation}
The generalized Laguere polynomials $L_n^{(\ell)}(r)$ have the following form 
\begin{equation}
L_n^{(\ell)}(r)= \frac{r^{-\ell} e^r}{n!} \frac{d^n}{dr^n} \big( e^{-r} r^{n+\ell}\big) = \sum\limits_{j=0}^n \frac{(-1)^j (n+\ell)!\, r^j}{j!(n-j)!(\ell+j)!},
\end{equation}
with normalization proportional to the Euler Gamma function
\begin{equation}\label{LNorm}
\int_0^\infty r^\ell e^{-r} L_n^{(\ell)}(r) L_m^{(\ell)}(r) dr= \frac{\Gamma(n+\ell+1)}{n!} \delta_{nm}.
\end{equation}
Since the gamma function and the energy \eqref{E_n} must be a positive we can restrict the quantum numbers in the following way:
\begin{equation}
n=0,1,2,...\,, \quad \ell=-n,-n+1,...,-1,0,1,2,...
\end{equation}

Switching back to the polar coordinates \eqref{PolCoord}, we obtain the final form of the wave function 
\begin{align}\label{Psi_nl}
\Psi_{n,\ell}(\rho,\varphi) = \frac{1}{\lambda}\sqrt{\frac{n!}{\pi(n+\ell)!}}\, e^{i\ell \varphi} \rho^\ell e^{-\frac{\rho^2}{2}} L_n^{(\ell)} \big(\rho^2\big).
\end{align}
Its normalization condition is given by
\begin{equation}
\lambda^2\int_0^\infty\!\!\! \int_0^{2\pi} \Psi_{n,\ell}^*(\rho,\varphi)\Psi_{n',\ell'}(\rho,\varphi)\, \rho\, d\rho\, d\varphi= \delta_{nn'} \delta_{\ell \ell'}.
\end{equation}

We are now ready to construct the Fock space of the system.

\subsection{Fock space}

We start by defining the following set of creation and annihilation operators\footnote{Note that $\partial_\mu^\dagger=-\partial_\mu$.}:
\begin{align} \label{defAB}
&a=-\frac{e^{i\varphi}}{2} \bigg(\rho +\partial_\rho +\frac{i}{\rho} \partial_\varphi \bigg), \quad  a^\dagger=-\frac{e^{-i\varphi}}{2} \bigg(\rho -\partial_\rho +\frac{i}{\rho} \partial_\varphi \bigg), \nonumber \\
&b=\frac{e^{-i\varphi}}{2} \bigg(\rho + \partial_\rho -\frac{i}{\rho} \partial_\varphi \bigg), \quad \,\, b^\dagger=\frac{e^{i\varphi}}{2} \bigg(\rho -\partial_\rho -\frac{i}{\rho} \partial_\varphi \bigg).
\end{align}
These operators obey the standard commutation relations:
\begin{equation} \label{commAB}
 [a,a^\dagger]=1 =[b,b^\dagger], \quad [a,b]=0.
\end{equation}
Their action on the wave function is given by:
\begin{align}
&a^\dagger \Psi_{n,\ell} = \sqrt{n+1}\, \Psi_{n+1,\ell-1}, \quad  b^\dagger \Psi_{n,\ell} = \sqrt{n+\ell+1}\, \Psi_{n,\ell+1}, \nonumber\\ 
&a \Psi_{n,\ell}= \sqrt{n}\, \Psi_{n-1,\ell+1}, \,\quad\qquad b \Psi_{n,\ell}= \sqrt{n+\ell\,}\, \Psi_{n,\ell-1} .
\end{align}
Therefore \( b^\dagger \) increases the quantum number \( \ell \) by one unit, without affecting \( n \). On the other hand, the operator \( a^\dagger \) raises \( n \) and lowers \( \ell \) by one unit. To simplify the notations, we introduce a shifted quantum number \( k = n + \ell \). The original wave function \( \Psi_{n,\ell} \) and the new one \( \Phi_{n,k}\) are related in a simple way:
\begin{equation}
\Psi_{n, k-n}= \Phi_{n,k} = \frac {(a^\dagger )^n (b^\dagger)^{k} } {\sqrt {n!k!} } \Phi_{0,0}, \quad \Psi_{0,0}=\Phi_{0,0}. 
\end{equation}
Now, the operators \( a \) and \( a^\dagger \) act only on \( n \), while the operators \( b \) and \( b^\dagger \) produce shift in \( k \):
\begin{align}
&a^\dagger \Phi_{n,k} = \sqrt{n+1}\, \Phi_{n+1,k}, \quad  b^\dagger \Phi_{n,k} = \sqrt{k+1}\, \Phi_{n,k+1}, \nonumber\\ 
&a \Phi_{n,k}= \sqrt{n}\, \Phi_{n-1,k}, \quad\qquad b \Phi_{n,k}= \sqrt{k}\, \Phi_{n,k-1} .
\end{align}
All these redefinitions does not affect the energy of the state \(\Phi_{n,k}\) so it matches the energy given by \eqref{E_n}. The Hamiltonian of our system takes the standard harmonic oscillator form
\begin{equation}\label{Ham}
H=\hbar\omega \bigg( a^\dagger a +\frac{1}{2} \bigg).
\end{equation}

Finally, the angular momentum \eqref{DL} can also be expressed in terms of the creation and annihilation operators: 
\begin{equation}
L_z=-\hbar\big( a^\dagger a -b^\dagger b \big), \quad L_z \Phi_{n,k}= \hbar \ell\, \Phi_{n,k}.
\end{equation}

This completes the quantization of the problem. We can now proceed to construct the TFD states of the system.

\section{Construction of TFD state}\label{sec3}

To construct the TFD state we make another copy of the Hilbert space and combine the two into left and right sectors. The two copies are considered independent of each other. In the \(\Phi_{n,k} \equiv |n,k \rangle\) basis, we can express the TFD state as \cite{chapman2019complexity, Avramov:2024kpt}:
\begin{align}\label{TFD}
|T\!F\!D \rangle&= \frac{1}{\sqrt{Z}}\sum\limits_{n=0}^\infty \sum\limits_{k=0}^\infty e^{-\frac{\beta E_n}{2}} |n,k \rangle_L |n,k \rangle_R  \nonumber \\ 
&=\frac{e^{-\frac{\beta\hbar \omega}{4}} }{\sqrt{Z}} \sum\limits_{n=0} ^\infty \sum\limits_{k=0}^\infty e^{-\frac{\beta\hbar\omega }{2}n} \frac{\big(a^\dag_L a^\dag_R\big)^n}{n!} \frac{\big(b^\dag_L b^\dag_R\big)^k}{k!}  |0,0 \rangle_L |0,0 \rangle_R \nonumber \\
&=\frac{e^{-\frac{\beta\hbar \omega}{4}} }{\sqrt{Z}} \exp \!\big(  e^{-\frac{\beta\hbar\omega }{2}} a^\dag_L a^\dag_R \big) \exp\! \big( b^\dag_L b^\dag_R \big) |0,0 \rangle_L |0,0 \rangle_R,
\end{align}
where $\beta$ is the inverse temperature, the energy \( E_{n} \) and the operators\footnote{The left and right operators commute: \([a_L, a_R^\dagger] = 0\) and \([b_L, b_R^\dagger] = 0\).} \( a_{L/R} \) and \( b_{L/R} \) are defined in \eqref{E_n} and \eqref{defAB}, respectively. The partition function $Z$ follows from the normalization condition of the TFD state
\begin{align}
1&=\langle T\!F\!D|T\!F\!D \rangle = \frac{e^{-\frac{\beta\hbar \omega}{2}}}{Z} \sum\limits_{n=0}^\infty \sum\limits_{n'=0}^\infty \sum\limits_{k=0}^\infty \sum\limits_{k'=0}^\infty  e^{-\frac{\beta\hbar\omega }{2}(n+n')} \delta_{nn'}\delta_{kk'}\delta_{nn'}\delta_{kk'}\nonumber\\
&= \frac{e^{-\frac{\beta\hbar \omega}{2}}}{Z} \sum\limits_{n=0}^\infty e^{-\beta\hbar\omega n} \sum\limits_{k=0}^\infty 1= \frac{e^{-\frac{\beta\hbar \omega}{2}}}{{Z\big( 1-e^{-\beta\hbar\omega} \big)}}   \big( 1 + \zeta(0) \big) =\frac{e^{-\frac{\beta\hbar\omega}{2}}}{2Z\big( 1-e^{-\beta\hbar\omega} \big)},
\end{align}
hence, the partition function is equal to half of the partition function for the harmonic oscillator:
\begin{equation}\label{PartF}
Z= \frac{e^{-\frac{\beta\hbar\omega}{2}}}{2\big( 1-e^{-\beta\hbar\omega} \big)} =\frac{1}{4  \sinh \frac{\beta\hbar\omega}{2}}.
\end{equation}
Note that we used the Riemann zeta function\footnote{Here we use   $\sum\limits_{k=1}^\infty 1=\zeta(0)=-1/2$.} for regularization of the sum.
Now we can write the normalized TFD state in the form:
\begin{align}\label{TFD1}
|T\!F\!D \rangle &= \sqrt{2 (1-e^{-\beta\hbar\omega})} \exp \!\big(  e^{-\frac{\beta\hbar\omega }{2}} a^\dag_L a^\dag_R \big) \exp\! \big( b^\dag_L b^\dag_R \big) |0,0 \rangle_L |0,0 \rangle_R \nonumber\\
&=\sqrt{2} \exp\big[\alpha\big(a_L^\dagger a_R^\dagger -a_L a_R \big)\big] \exp\! \big( b^\dag_L b^\dag_R \big) |0,0 \rangle_L |0,0 \rangle_R,
\end{align}
with parameter $\alpha$ defined by
\begin{equation}
\tanh\alpha=e^{-\frac{\beta\hbar\omega}{2}}.
\end{equation}
The calculation between the first and the second line of \eqref{TFD1} is related to the unitary decomposition of the TFD state presented in \cite{chapman2019complexity}. Since our TFD state is only $\alpha$ dependent, it is convenient to introduce the notation \(|T\!F\!D\rangle \equiv |\alpha\rangle\).  

The time-dependent TFD state follows by acting with the evolution operator \eqref{Ham} on \eqref{TFD}:
\begin{align}\label{TFD(t)}
|\alpha,t \rangle &= \frac{1}{\sqrt{Z}} \sum\limits_{n=0}^\infty \sum\limits_{k=0}^\infty  e^{-\frac{\beta E_n}{2}} e^{-\frac{i E_n}{\hbar} t} |n,k \rangle_L |n,k \rangle_R \nonumber \\
&=\frac{e^{-\frac{\beta\hbar \omega}{4}}  e^{-\frac{i\omega t}{2}} }{\sqrt{Z}} \exp \!\Big(  e^{-\frac{\beta\hbar\omega }{2}}  e^{-i \omega t} a^\dag_L a^\dag_R \Big) \exp\! \big( b^\dag_L b^\dag_R \big) |0,0 \rangle_L |0,0 \rangle_R \nonumber\\
&=\sqrt{2}\exp\big[\alpha\big( e^{-i\omega t}a_L^\dagger a_R^\dagger -e^{i\omega t}a_L a_R \big) \big] \exp\! \big( b^\dag_L b^\dag_R \big) |0,0 \rangle_L |0,0 \rangle_R.  
\end{align}
We express the creation and the annihilation operators with the corresponding left/right position and momentum operators
\begin{align}
a_{L/R}=\sqrt{\frac{m\omega}{2\hbar}} \bigg( X_{1\,L/R} +i\frac{P_{1\,L/R}}{m\omega} \bigg), \quad
b_{L/R}=\sqrt{\frac{m\omega}{2\hbar}} \bigg( X_{2\,L/R} +i\frac{P_{2\,L/R}}{m\omega} \bigg),
\end{align}
which allows us to write
\begin{equation}\label{a->L/R}
\alpha \big(a_L^\dagger a_R^\dagger -a_L a_R \big) = -\frac{i\alpha}{\hbar} \big( X_{1L} P_{1R} +X_{1R} P_{1L} \big),
\end{equation}
and consequently
\begin{equation}
b^\dagger_L b^\dagger_R = \frac{1}{2\hbar} \bigg( m\omega X_{2L}X_{2R} -\frac{P_{2L}P_{2R}}{m\omega} \bigg) -\frac{i}{2\hbar} \big( X_{2L}P_{2R} +X_{2R}P_{2L} \big).
\end{equation}
Finally, it is convenient to pass form $L/R$ basis to a new $\pm$ basis defined by:
\begin{equation}
X_{i\pm} =\frac{1}{\sqrt{2}} \big(X_{iL} \pm X_{iR} \big), \qquad P_{i\pm} =\frac{1}{\sqrt{2}} \big(P_{iL} \pm P_{iR} \big).  
\end{equation}
With these operator one can rewrite \eqref{a->L/R} as
\begin{align}
&\alpha\big(a_L^\dagger a_R^\dagger -a_L a_R \big) =-\frac{i\alpha}{\hbar} \big( X_{1+}P_{1+} -X_{1-}P_{1-} \big) 
=-i\alpha \hat K_+ +i\alpha \hat K_-,
\end{align}
where, using $[X_+,P_+]=[X_-,P_-]=i\hbar$, the operators $\hat K_\pm$ can be written by
\begin{equation}\label{K}
    \hat K_\pm=\frac{1}{2\hbar} \big( X_{1\pm}P_{1\pm} +P_{1\pm}X_{1\pm} \big).
\end{equation}
Consequently, we also have $b^\dagger_L b^\dagger_R =\hat A-i\hat B$, where the operators $\hat A$ and $\hat B$ are:
\begin{equation}\label{A,B,oper}
    \hat A=\frac{1}{4\hbar} \Big[ m\omega\big( X_{2+}^2 -X_{2-}^2 \big) -\frac{1}{m\omega} \big( P_{2+}^2 -P_{2-}^2 \big) \Big] ,\quad \hat B=\frac{1}{2\hbar} \big( X_{2+} P_{2+} -X_{2-} P_{2-} \big).
\end{equation}
Hence, our time-independent TFD state \eqref{TFD1} becomes
\begin{align}\label{TFDVac}
|\alpha\rangle=\sqrt{2}\,e^{-i\alpha \hat K_+} |0_{1+}\rangle \otimes e^{i\alpha \hat K_-} |0_{1-}\rangle \otimes  e^{\hat A-i\hat B} |0_2\rangle.
\end{align}

Similarly, we can define time-dependent operators 
\begin{equation}
e^{-i\omega t}a_L^\dagger a_R^\dagger -e^{i\omega t}a_L a_R
=-i\hat{K}_{+}(t)+i\hat{K}_{-}(t),
\end{equation}
where one has
\begin{align}\label{Kt}
\hat K_{\pm}(t)&=\frac{1}{2\hbar}\cos\omega t \big( X_{1\pm}P_{1\pm} +P_{1\pm}X_{1\pm} \big) +\frac{1}{2\hbar} \sin\omega t \bigg( m\omega X_{1\pm}^2 -\frac{P_{1\pm}^2}{m\omega} \bigg).
\end{align}
Note that at $t=0$ the operators $\hat K_\pm(0)= \hat K_\pm$ and $\big[ \hat K_+(t),\hat K_-(t) \big] =0$.
With these definitions, the time-dependent TFD state \eqref{TFD(t)} becomes
\begin{equation}\label{TFDt}
|\alpha,t\rangle=\sqrt{2}\, e^{-i\alpha \hat{K}_{+}(t)} |0_{1+}\rangle \otimes e^{i\alpha \hat{K}_-(t)} |0_{1-}\rangle \otimes e^{\hat A-i\hat B} |0_2\rangle.
\end{equation}

We can now proceed to determine the thermal covariance matrix of the system.

\section{Covariance matrix}\label{sec4}

Here we compute the thermal covariance matrix matrix of the system in both time-independent and time-dependent cases, following \cite{chapman2019complexity}.
\subsection{Covariance matrix for the time-independent TFD state}

It will be convenient to introduce the following vector operator
\begin{equation}
\vec\xi  =\vec\xi_{1+} \oplus\, \vec\xi_{1-} \oplus\, \vec\xi_{2+} \oplus\, \vec\xi_{2-} =(X_{1+},P_{1+},X_{1-},P_{1-},X_{2+},P_{2+},X_{2-},P_{2-})^T \equiv (\xi^r),
\end{equation}
where $\vec\xi_{i\pm}=(X_{i\pm},P_{i\pm})^T$.
The elements of the vacuum covariance matrix $O_0^{rs}\, (r,s=1,...,8)$ are
\begin{equation}
O_0^{rs} =\langle 0|\xi^r \xi^s |0\rangle = \frac{1}{2}\langle 0|\big\{ \xi^r, \xi^s \big\} |0 \rangle +\frac{1}{2} \langle 0| \big[\xi^r, \xi^s\big] |0 \rangle = \frac{\hbar}{2}\big( G^{rs}_0 +i\Omega^{rs}_0 \big),
\end{equation}
where the matrices $G_0$ and $\Omega_0$ are $8\times 8$ block-diagonal:
\begin{equation}\label{vacCovM}
G_0= \left(\!\!
\begin{array}{cccc}
\tilde G_0 & 0 & 0 & 0 \\
0 & \tilde G_0 & 0 & 0 \\
0 & 0 & \tilde G_0 & 0 \\
0 & 0 & 0 & \tilde G_0 \\
\end{array}
\!\!\right)\!, \quad  \Omega_0 = \left(\!\!
\begin{array}{cccc}
\tilde \Omega_0 & 0 & 0 & 0 \\
0 &\tilde  \Omega_0 & 0 & 0 \\
0 & 0 & \tilde \Omega_0 & 0 \\
0 & 0 & 0 & \tilde \Omega_0 \\
\end{array}
\!\!\right)\!.
\end{equation}
Their entries are the following $2\times 2$ matrices:
\begin{equation}
\tilde G_0= \left(\!\!
\begin{array}{cc}
\frac{1}{m\omega}  & 0 \\[5 pt]
0 & m\omega \\
\end{array}
\!\!\right)\!, \quad
\tilde \Omega_0 = \left(\!\!
\begin{array}{cc}
	0 & 1 \\
	-1 & 0  \\
\end{array}
\!\!\right)\!.
\end{equation}

The unitary operators $\hat U_\pm =e^{\mp i \alpha \hat K_\pm}$ and the operator $e^{\hat A-i\hat B}$ act on the vacuum to create our TFD state \eqref{TFDVac}:
\begin{equation}
|\alpha\rangle = \sqrt{2}\,\hat U_+ |0_{1+}\rangle \otimes \hat U_- |0_{1-}\rangle \otimes e^{\hat A-i\hat B} |0_2\rangle, 
\end{equation}
where $\hat K_{\pm}$ are Hermitian generators defined in \eqref{K}. Here we introduce the matrix representations  $\mathcal U_{\pm}$ and $\mathcal K_{\pm}$ of the operators $\hat U_\pm$ and $\hat K_\pm$\footnote{The indices $a,b=1,2$, thus $\xi^1_{1\pm}=X_{1\pm}$ and $\xi^2_{1\pm}=P_{1\pm}$.} by the following relations:
\begin{equation}\label{MaricesUK}
\hat U_{\pm}^\dagger \xi_{1\pm}^a \hat U_{\pm} =\mathcal U_{\pm,b\,}^{\,\,\,a} \xi_{1\pm}^b, \quad 
[i\hat K_\pm,\xi_{1\pm}^a] =\big(\mathcal K_{\pm}. \vec\xi_{1\pm} \big)^a =\mathcal K_{\pm,b\,}^{\,\,\,a} \xi_{1\pm}^b.
\end{equation}
It is easy to show that $\mathcal U_\pm=e^{\pm\alpha \mathcal K_\pm}$, i.e.
\begin{align}
\hat U_{\pm\,}^\dagger \xi_{1\pm}^a \hat U_{\pm}&= e^{\pm i\alpha \hat K_{\pm\,}} \xi_{1\pm\,}^a e^{\mp i\alpha \hat K_{\pm}} =\sum_{n=0}^\infty \frac{(\pm\alpha)^n}{n!} \big[ i\hat K_{\pm},\xi_{1\pm}^a \big]_{(n)} \nonumber\\  
& =\sum_{n=0}^\infty \frac{(\pm\alpha)^n}{n!}  \big(\mathcal K_{\pm}^n .\vec\xi_{1\pm} \big)^a  =\Big( e^{\pm\alpha\mathcal K_\pm} .\vec\xi_{1\pm} \Big)^{\!a} =\Big( \mathcal U_{\pm} .\vec\xi_{1\pm} \Big)^{\!a} =\mathcal U_{\pm,b\,}^{\,\,\,a} \xi_{1\pm}^b,
\end{align}
where  $\big[ i\hat K_+,\xi^a_{1+} \big]_{(n)}$ denotes the $n$-th nested commutator\footnote{For example $\big[ i\hat K_+,\xi^a_{1+} \big]_{(2)} \equiv \big[i\hat K_+,\big[ i\hat K_+,\xi^a_{1+}\big] \big] =\big[i\hat K_+,\big(\mathcal K_+ .\vec\xi_{1+} \big)^a \big]= \big( \mathcal K_+.\mathcal K_+.\vec\xi_{1+}\big)^a =\big( \mathcal K_+^2.\vec\xi_{1+}\big)^a$.}. The commutators between $\hat K_\pm$ and $\xi_{1\pm}^a$ are written by
\begin{equation}
[i\hat K_\pm,X_{1\pm}]=   X_{1\pm}, \quad  [i\hat K_{\pm},P_{1\pm}]= - P_{1\pm}, 
\end{equation}
which set the form of the matrices $\mathcal K_+=\mathcal K_-=\mathcal K$:
\begin{equation}
\mathcal K =\left(\!\!
\begin{array}{cc}
	1 \!& 0  \\
	0 \!& -1 
\end{array}
\!\!\right)\!.
\end{equation}
Exponentiating $\mathcal K$ yields $\mathcal U_\pm$,
\begin{equation}
\mathcal U_\pm =e^{\pm\alpha \mathcal K}= \mathbb{1}\, \cosh\alpha \pm \mathcal K\, \sinh \alpha =\left(\!\!
\begin{array}{cc}
e^{\pm \alpha} \!& 0 \\
0 \!& e^{\mp\alpha} \\
\end{array}
\!\!\right)\!.	
\end{equation}

The full TFD covariance matrix can be constructed by $O^{rs}=\langle \alpha|\xi^r \xi^s |\alpha\rangle$. 
Here, we demonstrate how to calculate the upper-left $2\times 2$ block:
\begin{align}
\tilde O_{1+}^{ab}&=\langle \alpha|\xi_{1+}^a \xi_{1+}^b |\alpha\rangle =2\langle 0_2| e^{\hat A^\dagger +i\hat B^\dagger} e^{\hat A-i\hat B} |0_{2}\rangle \langle 0_{1-}| \hat U^\dagger_{-} \hat U_{-} |0_{1-}\rangle \langle 0_{1+}|\hat U^\dagger_+ \xi_{1+}^{a}\xi_{1+}^{b} \hat U_+  |0_{1+}\rangle  \nonumber \\
&= \mathcal U_{+,c}^{\,\,\,a} \langle 0_{1+}| \xi_{1+}^{c}\xi_{1+}^{d} |0_{1+}\rangle\, \mathcal U_{+,d}^{\,\,\,b} = \mathcal U_{+,c\,}^{\,\,\,a} \tilde O_{0}^{cd\,} \mathcal U_{+,d}^{\,\,\,b} \nonumber\\
&=\frac{\hbar}{2} \Big( \mathcal U_{+,c\,}^{\,\,\,a} \tilde G_{0}^{cd\,} \mathcal U_{+,d}^{\,\,\,b}  +i\,\mathcal U_{+,c\,}^{\,\,\,a} \tilde \Omega_{0}^{cd\,} \mathcal U_{+,d}^{\,\,\,b} \Big) =\frac{\hbar}{2} \Big( \tilde G_{1+}^{ab}  +i \tilde \Omega_{1+}^{ab}  \Big),
\end{align}
where we used 
\begin{equation}
\langle 0_2| e^{\hat A^\dagger +i\hat B^\dagger} e^{\hat A-i\hat B} |0_{2}\rangle =\frac{1}{2}, \quad \langle 0_{1-}| \hat U^\dagger_{-} \hat U_{-} |0_{1-}\rangle=1.
\end{equation}
Similar calculations can be done for the block $\tilde O^{ab}_{1-} =\langle \alpha|\xi_{1-}^a \xi_{1-}^b |\alpha\rangle$. Therefore, we can write the explicit form of $\tilde G_{1\pm}$ and $\tilde \Omega_{1\pm}$ simply using matrix multiplication:
\begin{equation}\label{G1pm}
\tilde G_{1\pm}=\mathcal U_{\pm}.\tilde G_0.\,\mathcal U_{\pm}^T= \left(\!\!
\begin{array}{cc}
\dfrac{e^{\pm2\alpha}}{m\omega} \!& 0 \\[5pt]
	0 \!& m\omega e^{\mp2\alpha}
\end{array}
\!\!\right)\!,\quad
\tilde \Omega_{1\pm}=\mathcal U_{\pm}.\tilde \Omega_0.\,\mathcal U_{\pm}^T =\tilde \Omega_0. 
\end{equation}
The calculations for the remaining part of the covariance matrix are slightly different:
\begin{align}
&\tilde O_{2+}^{ab}=\langle \alpha|\xi_{2+}^a \xi_{2+}^b |\alpha\rangle =2\langle 0_{1-}| \hat U^\dagger_{-} \hat U_{-} |0_{1-}\rangle \langle 0_{1+}|\hat U^\dagger_+ \hat U_+  |0_{1+}\rangle  \langle 0_2| e^{\hat A^\dagger +i\hat B^\dagger} \xi_{2+}^{a}\xi_{2+}^{b} e^{\hat A-i\hat B} |0_{2}\rangle \nonumber \\
&=\langle 0_2| e^{\hat A^\dagger +i\hat B^\dagger} \{\xi_{2+}^{a},\xi_{2+}^{b} \} e^{\hat A-i\hat B} |0_{2}\rangle + \langle 0_2| e^{\hat A^\dagger +i\hat B^\dagger} [\xi_{2+}^{a},\xi_{2+}^{b} ] e^{\hat A-i\hat B} |0_{2}\rangle = \frac{\hbar}{2}\big( \tilde G^{ab}_{2+} +i \tilde\Omega^{ab}_{2+} \big).
\end{align}
For example, we can show how to calculate the matrix element $\tilde G_{2+}^{11}$. To do that, we switch to the $b$ and $b^\dagger$ operators and use the generalized zeta function $\zeta(s,a)=\sum\limits_{k=0}^\infty \dfrac{1}{(k+a)^s}$, hence:
\begin{align}
\tilde G_{2+}^{11}&=\frac{2}{\hbar} \langle 0_2| e^{\hat A^\dagger +i\hat B^\dagger} \{X_{2+},X_{2+} \} e^{\hat A-i\hat B} |0_{2}\rangle =\frac{4}{\hbar} \langle 0_2| e^{\hat A^\dagger +i\hat B^\dagger} X_{2+} X_{2+} e^{\hat A-i\hat B} |0_{2}\rangle \nonumber \\
&=\frac{4}{\hbar} {}_R\langle 0| {}_L\langle 0| e^{b_Lb_R} \frac{\hbar}{2m\omega} \big( X_{2L}+X_{2R}\big) \big( X_{2L}+X_{2R}\big) e^{b_L^\dagger b_R^\dagger} |0\rangle_L |0\rangle_R \nonumber \\
&=\frac{2}{m\omega} \sum_{k=0}^{\infty} \sum_{k'=0}^{\infty} {}_R\langle k'|{}_L\langle k'| \big( X_{2L}X_{2L} +X_{2R}X_{2R} \big) |k\rangle_L |k\rangle_R \nonumber \\
&=\frac{1}{m\omega} \sum_{k=0}^{\infty} \sum_{k'=0}^{\infty} {}_R\langle k'|{}_L\langle k'| \big( b_L^\dagger b_L + b_L  b_L^\dagger + b_R^\dagger b_R + b_R  b_R^\dagger \big) |k\rangle_L |k\rangle_R \nonumber \\
&=\frac{1}{m\omega} \sum_{k=0}^{\infty} \sum_{k'=0}^{\infty} {}_R\langle k'|{}_L\langle k'| \big( 4k +2 \big) |k\rangle_L |k\rangle_R =\frac{1}{m\omega} \sum_{k=0}^{\infty} \sum_{k'=0}^{\infty} \big( 4k +2 \big) \delta_{k'k}\delta_{k'k} \nonumber \\
&=\frac{4}{m\omega} \sum_{k=0}^{\infty} k +\frac{1}{2} =\frac{4}{m\omega} \zeta\bigg(\!\!-\!1,\frac{1}{2} \bigg) =\frac{4}{m\omega} \frac{1}{24} =\frac{1}{6m\omega}. 
\end{align}
After some lengthy but similar calculations, we find the matrices $\tilde G_{2+} =\tilde G_{2-}=\tilde G_2$ and $\Omega_{2\pm}$:
\begin{equation}\label{CovM2}
\tilde G_{2} =\frac{1}{6} \tilde G_0 =\frac{1}{6} \left(\!\!
\begin{array}{cc}
	\frac{1}{m\omega} \!& 0 \\
	0 \!& m\omega
\end{array}
\!\!\right)\!, \quad \tilde\Omega_{2\pm} =\tilde\Omega_0.
\end{equation}

\subsection{Covariance matrix for the time-dependent TFD state}

The derivation of the covariance matrix for the time-dependent case follows the same steps. We create the time-dependent TDF state \eqref{TFDt} by acting on the vacuum with the unitary operators $\hat U_{\pm}(t)=e^{\mp i \alpha \hat K_{\pm}(t)}$ and the operator $e^{\hat A-i\hat B}$:
\begin{equation}
|\alpha,t\rangle = \sqrt{2}\, \hat U_+(t) |0_{1+}\rangle \otimes \hat U_-(t) |0_{1-}\rangle \otimes e^{\hat A-i\hat B} |0_2\rangle . 
\end{equation}
Next we introduce the matrices $\mathcal U_{\pm}(t)$ and $\mathcal K_{\pm}(t)$ in the same way as in \eqref{MaricesUK}:
\begin{equation}
\hat U_{\pm}^\dagger(t)\, \xi_{1\pm}^a \hat U_{\pm}(t) =\mathcal U_{\pm,b}^{\,\,\,a} (t)\, \xi_{1\pm}^b, \quad  [i\hat K_\pm(t),\xi_{1\pm}^a] =\mathcal K_{\pm,b}^{\,\,\,a}(t)\, \xi_{1\pm}^b.
\end{equation}
The commutation relations between $\hat K_\pm(t)$ and $\xi_{1\pm}^a$ are given by:
\begin{align}
[i\hat K_\pm(t),X_{1\pm}] &=  \cos\omega t\, X_{1\pm} -\frac{\sin\omega t}{m \omega} P_{1\pm}, \nonumber \\
[i\hat K_\pm(t),P_{1\pm}] &=  -\cos\omega t\, P_{1\pm} -m\omega \sin\omega t\, X_{1\pm},
\end{align}
which lead to the explicit form of the matrices $\mathcal K_+(t)=\mathcal K_-(t)=\mathcal K(t)$:
\begin{equation}
\mathcal K(t) =\left(\!\!
\begin{array}{cc}
\cos\omega t & -\dfrac{\sin\omega t}{m\omega} \\[5pt]
-m\omega \sin\omega t
& -\cos\omega t
\end{array}
\!\right)\!.
\end{equation}
Therefore the matrices $\mathcal U_\pm(t)$ yields
\begin{equation}
\mathcal U_\pm(t) =e^{\pm\alpha \mathcal K(t)}\!= \mathbb{1}\! \cosh\alpha \pm \mathcal K(t) \sinh \alpha =\!\! \left(\!\!\!
\begin{array}{cc}
\cosh\alpha \pm \sinh\alpha \cos\omega t \!\! &\!\! \mp \dfrac{ \sinh \alpha \sin\omega t}{m\omega} \\[5pt]
\mp m\omega \sinh\alpha \sin\omega t \!\!&\!\! \cosh\alpha \mp \sinh\alpha \cos\omega t \\
\end{array}
\!\!\!\right)\!\!.	
\end{equation}
The time-dependent covariance matrix is defined by $O^{rs}(t)\!=\!\langle \alpha,t|\xi^r \xi^s |\alpha,t\rangle$. The first part $\tilde G_{1\pm}(t)$ and $\tilde \Omega_{1\pm}(t)$ can be calculated using matrix multiplication:
\begin{equation}\label{G1pmt}
\tilde G_{1\pm}(t)\! =\mathcal U_\pm(t).\tilde G_0.\,\mathcal U_\pm^T(t) =\!\left(\!\!
\begin{array}{cc}
\dfrac{ \cosh 2\alpha \pm \sinh 2\alpha \cos\omega t}{m\omega}  \!&\! \mp \sinh 2\alpha \sin\omega t \\[5pt]
\mp\, \sinh 2\alpha \sin\omega t \!&\! m\omega \big( \cosh 2\alpha \mp \sinh 2\alpha \cos\omega t \big) \\
\end{array}
\!\!\!\right)\!\!,
\end{equation}
and
\begin{equation}
\tilde\Omega_{1\pm}(t)=\mathcal U_\pm(t).\tilde\Omega_0.\,\mathcal U_{\pm}^T(t)= \tilde\Omega_0.
\end{equation}
At $t=0$ the matrix $\tilde G_{1\pm}(0)= \tilde G_{1\pm}$ \eqref{G1pm}.
The other part of the covariance matrix is time-independent $\tilde G_{2\pm}(t)= \tilde G_{2}$ and $\tilde\Omega_{2\pm}(t) =\tilde \Omega_0$, which coincides with \eqref{CovM2}. As a result of our computations above we can write the full $8\times 8$ covariance matrix
\begin{equation}\label{covGt}
G(t)= \left(\!\!
\begin{array}{cccc}
\tilde G_{1+}(t) & 0 & 0 & 0 \\
0 & \tilde G_{1-}(t) & 0 & 0 \\
0 & 0 & \tilde G_2 & 0 \\
0 & 0 & 0 & \tilde G_2 \\
\end{array}
\!\!\right)\!,
\end{equation}
with $\tilde G_{1\pm}(t)$ and $\tilde G_2$ defined in \eqref{G1pmt} and \eqref{CovM2}.
Finally, we can focus on finding the complexity of our system.

\section{Complexity}\label{sec5}

In this section, we compute the relative covariance matrix and its eigenvalues, which allows us to analyze its behavior in different regimes of the parameters.

\subsection{Relative covariance matrix and complexity}

A natural choice of the target state is the full $8\times 8$ time-dependent covariance matrix \eqref{covGt}, while the reference state can be represented by the 
vacuum covariance matrix \eqref{vacCovM} with a reference frequency $\omega_{R}\neq \omega$:
\begin{equation}
G_R=\left(\!\!
\begin{array}{cccc}
\tilde G_R & 0 & 0 & 0 \\
0 & \tilde G_R & 0 & 0 \\
0 & 0 & \tilde G_R & 0 \\
0 & 0 & 0 & \tilde G_R \\
\end{array}
\!\!\right)\!, \quad   \tilde G_R=\left(\!\!
\begin{array}{cc}
\frac{1}{m\omega_{R}} & 0\\[5pt]
0 & m\omega_{R}
\end{array}
\!\!\right)\!.
\end{equation}
By definition the relative covariance matrix $\Delta(t)$ is
\begin{equation}\label{eqRelCovMatr}
\Delta(t)=G(t).G_R^{-1}=\left(\!
\begin{array}{cccc}
\tilde\Delta_{1+}(t) & 0 & 0 & 0 \\
0 & \tilde\Delta_{1-}(t) & 0 & 0 \\
0 & 0 & \tilde\Delta_{2} & 0 \\
0 & 0 & 0 & \tilde\Delta_{2} \\
\end{array}
\!\right)\!,
\end{equation}
where the $2\times 2$ matrices $\tilde\Delta_{1\pm}(t)$ and $\tilde\Delta_2$ assume the following form:
\begin{equation}
\tilde \Delta_{1\pm}(t)=\tilde G_{1\pm}(t).\tilde G_R^{-1}= \left(\!
\begin{array}{cc}
\dfrac{\omega_{R}}{\omega} \big( {\rm{ch}}2\alpha \pm {\rm{sh}}2\alpha \cos\omega t \big) & \mp\, \dfrac{{\rm{sh}} 2\alpha \sin\omega t}{m\omega_{R}} \\[5pt]
\mp\, m\omega_{R}\, {\rm{sh}}2\alpha \sin\omega t & \dfrac{\omega}{\omega_{R}} \big( {\rm{ch}}2\alpha \mp {\rm{sh}}2\alpha \cos \omega t \big) \\ \end{array}
\!\!\right)\!,	
\end{equation}
and 
\begin{equation}
\tilde \Delta_2=\tilde G_2.\tilde G_R^{-1}=\left(\!
\begin{array}{cc}
\dfrac{\omega_R}{6\omega} & 0\\[5pt]
0 & \dfrac{\omega}{6\omega_R}
\end{array}
\!\!\right)\!.
\end{equation}

The eigenvalues of $\Delta(t)$ are all positive:
\begin{align}\label{eigenvalues}
&e_1=A_{+}-\sqrt{A_{+}^2-1}\,, \quad e_2=A_{+}+\sqrt{A_{+}^2-1}\,, \nonumber \\
&e_3=A_{-}-\sqrt{A_{-}^2-1}\,, \quad e_4=A_{-}+\sqrt{A_{-}^2-1}\,, \nonumber \\
&e_5=e_7=\frac{\omega_R}{6\omega}, \qquad \qquad\,\, e_6=e_8=\frac{\omega}{6\omega_R},
\end{align}
where $A_\pm$ are time-dependent functions written by:
\begin{equation}\label{AAA}
A_{\pm}=\frac{1}{2\omega_R\,\omega} \Big( \big(\omega_R^2+\omega^2\big) \cosh 2\alpha \pm \big(\omega_R^2-\omega^2\big) \sinh 2\alpha \cos \omega t \Big).
\end{equation}
One notes that the eigenvalues become time-independent if \(\omega_R = \omega\). Additionally, at zero temperature limit (\(\alpha \to 0\)), the eigenvalues are also time-independent. Knowledge of the eigenvalues allows us to write the relative matrix in a diagonal form \(\Delta = \mathrm{diag}(e_1, \ldots, e_8)\), which simplifies the sub-sequence calculations. For example, one notes that the matrix \(\sqrt{\Delta}\) serves as a linear map between the states \cite{chapman2019complexity}:
\begin{equation}
 \big( \sqrt{\Delta} \big)^2.G_R= \Delta.G_R= G(t).G_R^{-1}.G_R =G(t).
\end{equation}
It can be expressed in an exponential form $\sqrt{\Delta}=e^M$, with a diagonal generator $M$ given by $M=\ln \sqrt{\Delta} =\frac{1}{2}\ln \Delta$ $=\frac{1}{2}\, {\rm{diag}} \big( \ln e_1, ...,\ln e_8 \big)$. Hence, the complexity, which defines the geodesic distance between the reference state \(G_R\) and the target state \(G(t)\), can be calculated using the Frobenius norm of the generator $\mathcal C(t)=|\!|M|\!|=  \sqrt{{\rm{Tr}} M^2}$, i.e. 
\begin{align}\label{Complexity}
\mathcal C(t) &= \frac{1}{2} \sqrt{ {\rm{Tr}} (\ln \Delta)^2 } =\frac{1}{2} \sqrt{ \sum_{s=1}^8 \ln^2 e_s\,} = \sqrt{ \ln^26 +\ln^2\frac{\omega_R}{\omega} + \frac{1}{4}\sum_{s=1}^4 \ln^2 e_s\, }  .
\end{align}
Doe to \eqref{AAA}, the complexity \(\mathcal{C}(t)\) is a periodic function with a period $\mathcal T=\pi/\omega$.

\subsection{Complexity for $\omega_{R}=\omega$}

If we set the reference frequencies equal to the target frequency, \(\omega_R = \omega\), the complexity \eqref{Complexity} becomes time-independent and simplifies to
\begin{equation}\label{C_wR=w}
\mathcal C= \sqrt{ \ln^26 +4\alpha^2 \, } =\sqrt{ \ln^26 + 4\,{\rm{arcth}}^2 e^{-\frac{\beta\hbar\omega}{2}} \,}.
\end{equation}

At low temperature, \(\beta \hbar \omega = \frac{\hbar \omega}{kT} \gg 1\), the asymptotic expansion of \eqref{C_wR=w} and its limit are:
\begin{equation}
\mathcal C\approx \ln 6 +\frac{2 e^{-\beta\hbar\omega}}{\ln 6} , \quad \lim_{\beta\to\infty} \mathcal C= \ln 6\approx 1.792.
\end{equation}
In this case we observe that the temperature dependence of complexity is exponentially suppressed, hence at zero temperature $(\beta\to \infty)$, the complexity acquires positive lower bound.  

At high temperature, $\beta \hbar\omega =\dfrac{\hbar\omega}{kT}\ll 1$, complexity and its limit are given by:
\begin{equation}
\mathcal C \approx \ln\frac{4}{\beta\hbar\omega} +\frac{\ln^26}{2\ln\frac{4}{\beta\hbar\omega}}, \quad \lim_{\beta\to 0} \mathcal C= \infty.
\end{equation}
One notes that complexity has logarithmic dependence on temperature and hence at $\beta\to 0$ it diverges.

\subsection{Complexity for $\omega_{R}\neq\omega$}

In this subsection we consider different reference and target frequencies.

\subsubsection{Temperature analysis}

\begin{figure}[H]
\centering \hspace{-1.0cm}
\begin{subfigure}{0.4\textwidth}
\includegraphics[width=8.3cm,height=5.55cm]{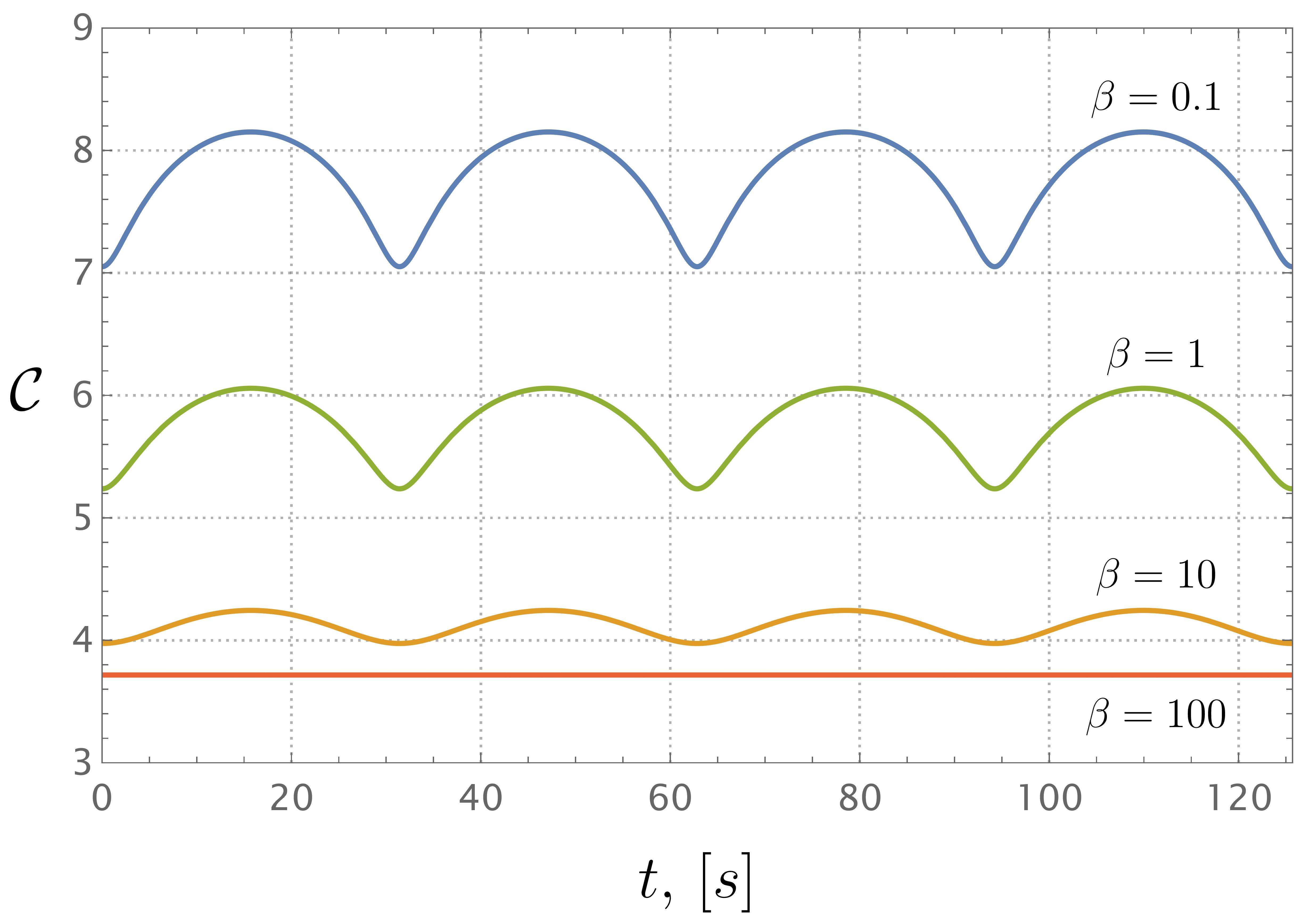}
\caption{Complexity $\mathcal C(t)$.}\label{Cc}
\end{subfigure}
\hspace{1.5 cm}
\begin{subfigure}{0.4\textwidth}
\includegraphics[width=8.3cm,height=5.5cm]{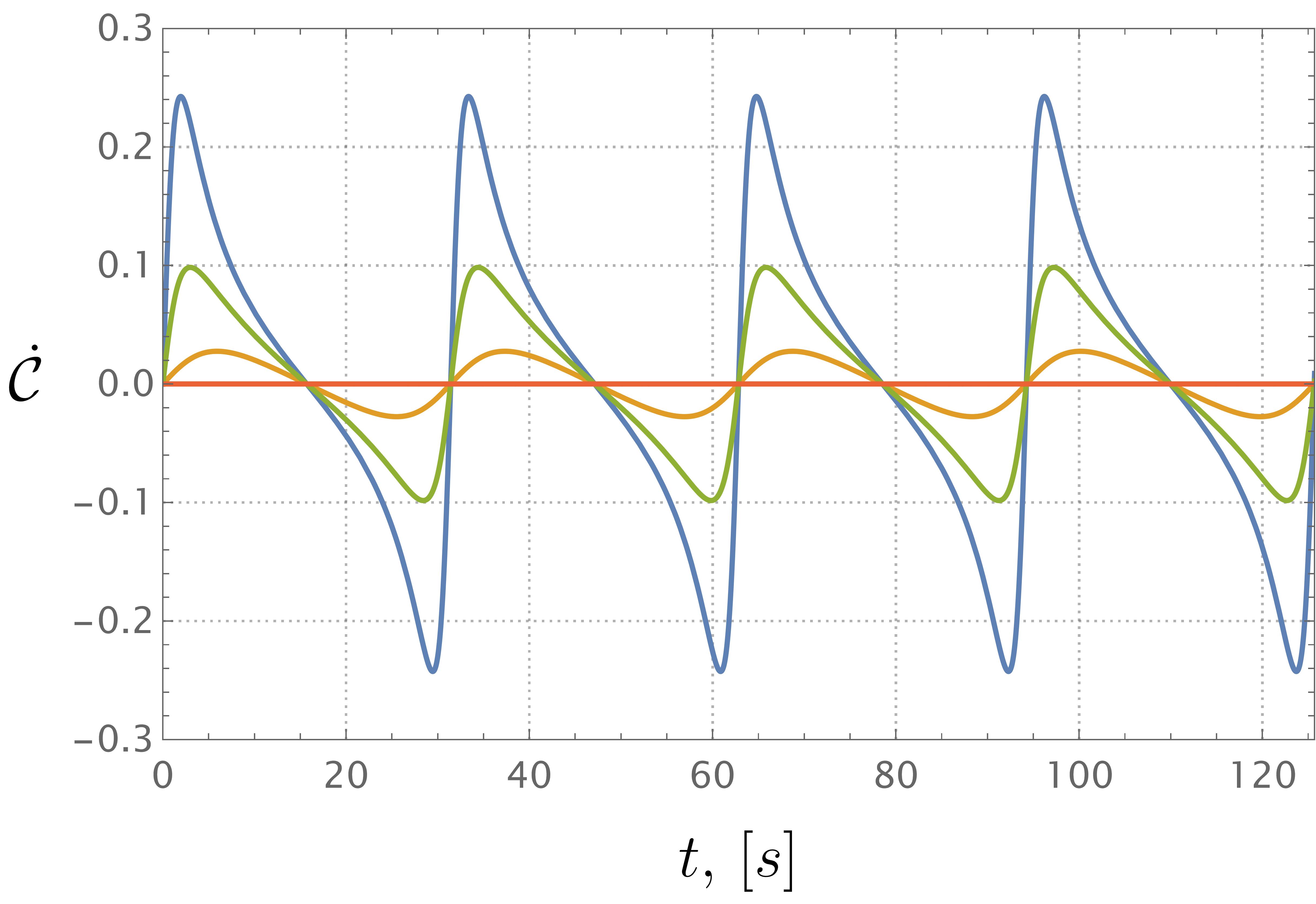}\vspace{-0.05cm}
\caption{Rate of complexity $\dot{\mathcal C}(t)$. }\label{Rate}
\end{subfigure}
 \hspace{0.7 cm}
\caption{ \textbf{(a)} Complexity $\mathcal C(t)$ at different temperatures.
\textbf{(b)}  The corresponding rate of  complexity $\dot{\mathcal C}(t)$. Here the blue curve corresponds to infinite temperature $\beta\to 0$. We choose: $\omega_R=1,\, \omega=0.1$ and $\hbar=1$.
}\label{figEnsembles}
\end{figure}

\begin{figure}[H]
\centering \hspace{-1.0cm}
\begin{subfigure}{0.4\textwidth}
\includegraphics[width=8.3cm,height=5.55cm]{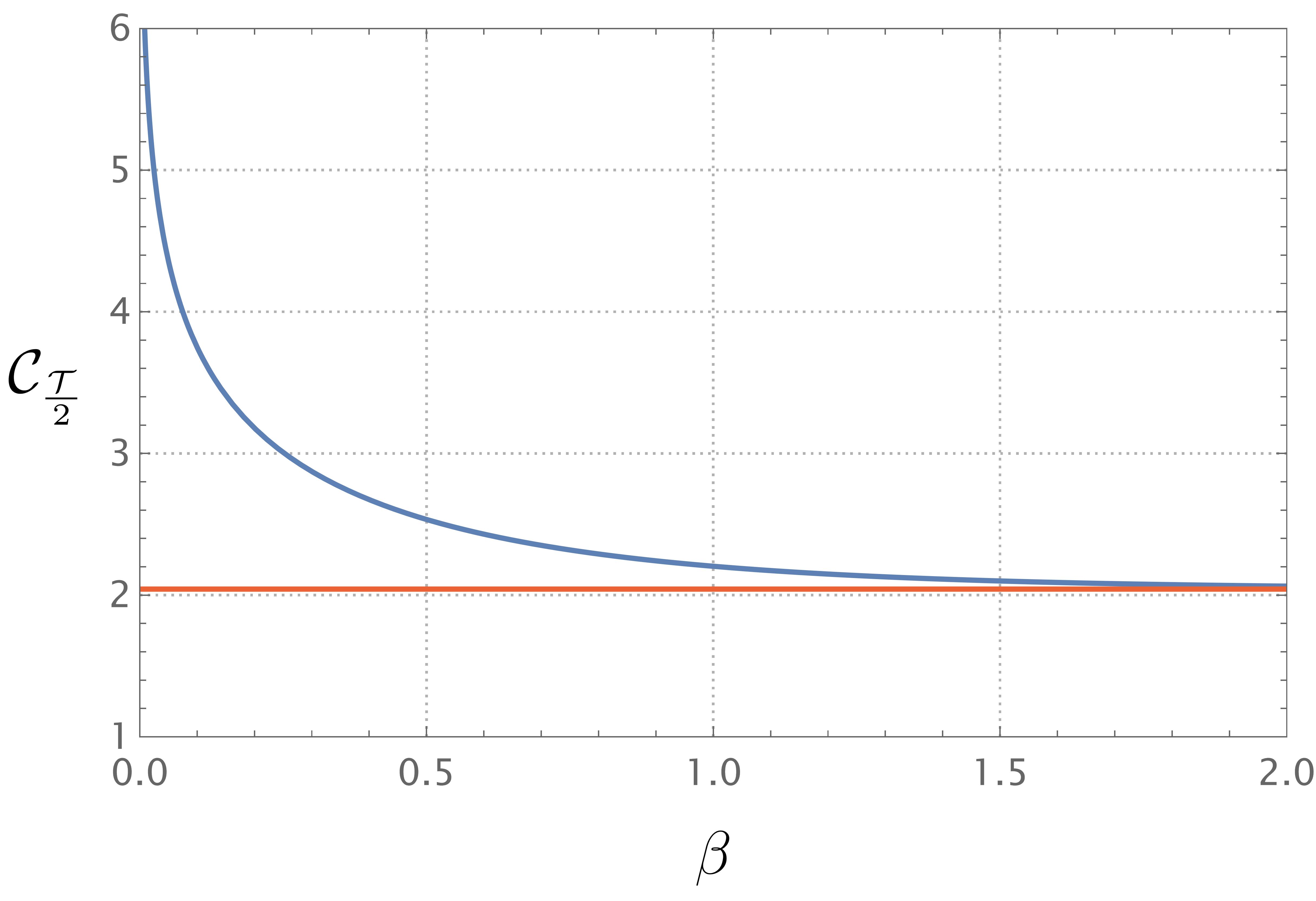}
\caption{Complexity $\mathcal C_{\frac{\mathcal{T}}{2}}(\beta)$.}\label{C(b)}
\end{subfigure}
\hspace{1.5 cm}
\begin{subfigure}{0.4\textwidth}
\includegraphics[width=8.3cm,height=5.5cm]{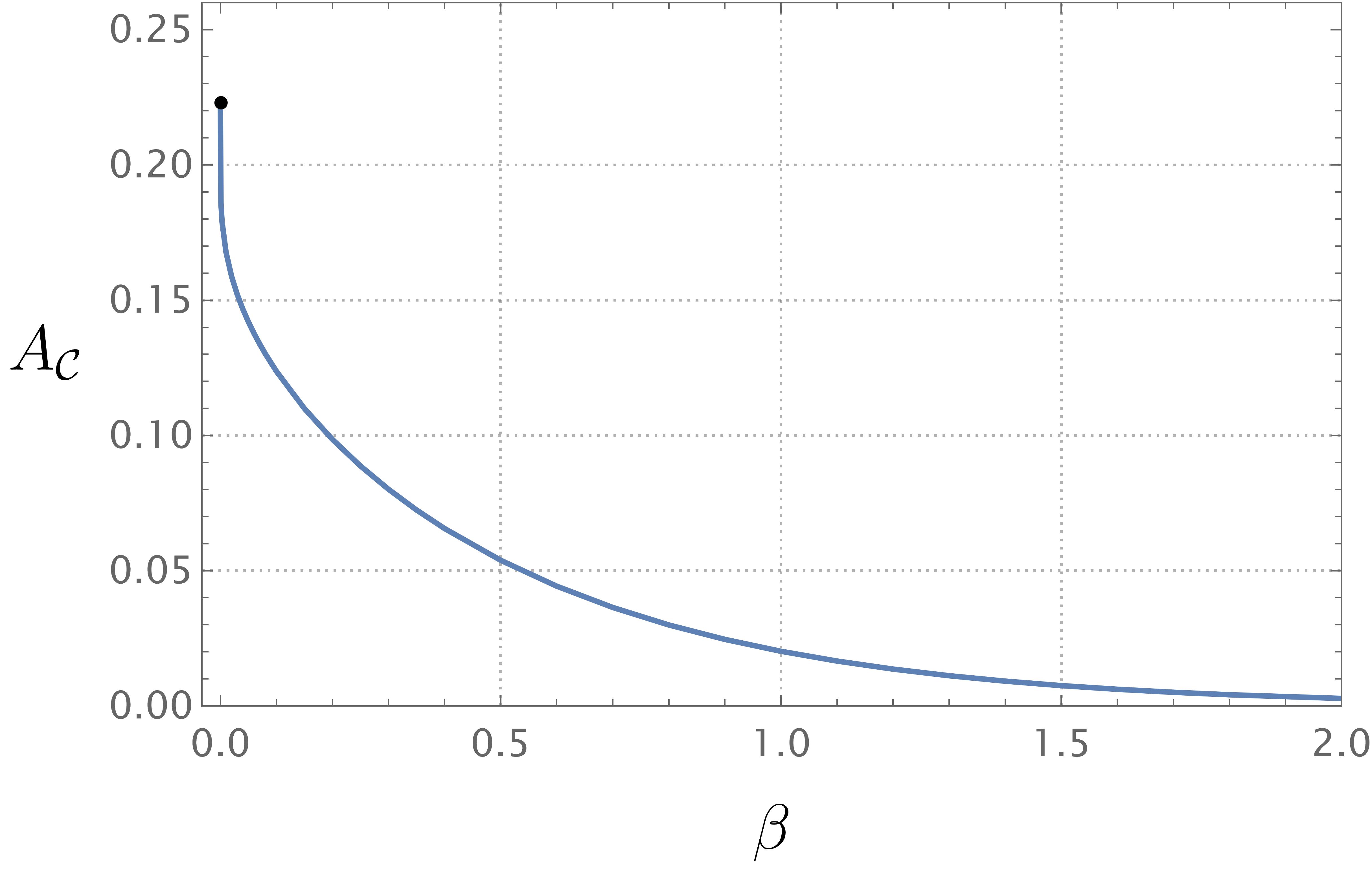}\vspace{-0.05cm}
\caption{Amplitude of oscillations $A_{\mathcal C}(\beta)$. }\label{Ac(b)}
\end{subfigure}
 \hspace{0.7 cm}
\caption{ \textbf{(a)} Complexity $\mathcal C_{\frac{\mathcal{T}}{2}}(\beta)$ as a function of $\beta$ at $t=\mathcal{T}/2$.
\textbf{(b)}  The amplitude of  complexity oscillations $A_{\mathcal C}(\beta)$. We choose: $\omega_R=1,\, \omega=2$ and $\hbar=1$.
}\label{figCb}
\end{figure}

At low temperature (\(\beta \hbar \omega \gg 1\)), the asymptotic expansion of \eqref{Complexity} is given by:
\begin{align}\label{red}
&\mathcal C(t)\approx \sqrt{ \ln^26 +2\ln^2\frac{\omega_{R}}{\omega} } + \frac{ 2e^{-\beta\hbar\omega} }{\sqrt{ \ln^26  + 2\ln^2 \frac{\omega_R}{\omega} }} \bigg(\! \cos^2\omega t +\frac{\omega_R^2+\omega^2}{\omega_R^2-\omega^2} \ln\frac{\omega_R}{\omega} \,\sin^2\omega t \bigg).
\end{align}
Once again, besides the first term, all other terms are exponentially  suppressed with respect to the temperature, hence at zero temperature $(\beta\to \infty)$ complexity saturates at a minimum value (the red line on Fig.\,\ref{Cc} and Fig.\,\ref{C(b)}):
\begin{equation}\label{minValOfC}
    \lim_{\beta\to\infty} \mathcal C(t)= \sqrt{ \ln^26 +2\ln^2\frac{\omega_{R}}{\omega} }.
\end{equation}

At high temperature (\(\beta \hbar \omega \ll 1\)), the asymptotic expansion of complexity and its limit are:
\begin{equation}\label{HighTt}
{\cal C}(t) \approx   \ln\frac{1}{\beta\hbar\omega} +\ln \frac{2\sqrt{(\omega_R^2+ \omega^2)^2 -(\omega_R^2-\omega^2)^2 \cos^2\omega t}}{\omega_{R\,} \omega}, \quad  \lim\limits_{\beta\to 0} \mathcal C(t)= \infty.
\end{equation}
We observe similar logarithmic behavior of complexity as in the previous subsection, hence $\mathcal C(t)$ diverges for $\beta\to 0$.
As evident from Fig.\,\ref{Cc} complexity has minima at $t=m \mathcal{T}$, and maxima at $t=\mathcal{T}/2+m\mathcal{T}$, where $m=0,1,2,...$ and $\mathcal{T} =\pi/\omega$ is the period of $\mathcal C(t)$. Since all maxima have the same value, we can define the local maximum of complexity as $\mathcal C(\mathcal{T}/2)\equiv \mathcal C_{\frac{\mathcal{T}}{2}}$. The later is function only on the temperature and the frequency. 

Due to the fact that complexity diverges at certain points as a function of temperature, it is useful to define a new quantity called the amplitude of complexity oscillations, which remains finite at all temperatures. This time-independent quantity is defined as the difference between the maximum and minimum values of complexity
\begin{equation}
A_{\mathcal{C}}= \mathcal{C}(\mathcal{T}/2) -\mathcal{C}(0).
\end{equation}
At low temperature $(\beta\hbar\omega  \gg 1)$ we can calculate the asymptotic behavior of $A_\mathcal{C}$ is (see \eqref{red}):
\begin{equation}\label{ALowT}
A_{\mathcal C} \approx \frac{2e^{-\beta\hbar\omega}}{\sqrt{ \ln^26 +2\ln^2\frac{\omega_R}{\omega}}} \bigg( \frac{\omega_R^2 +\omega^2}{\omega_R^2 -\omega^2} \ln\frac{\omega_R}{\omega} -1 \bigg), \quad  \lim_{\beta\to \infty} A_{\mathcal{C}}= 0.
\end{equation}
We note that the amplitude of the oscillation is exponentially suppressed with respect to the temperature and vanishes at $(\beta\to \infty)$. At high temperature (\(\beta \hbar \omega \ll 1\)) one has:
\begin{equation}\label{AHighT}
A_{\mathcal C} \approx \ln \frac{\omega_R^2+\omega^2}{2\omega_R\, \omega } -\frac{\ln^2 \frac{\omega_R}{\omega} }{2\ln{\frac{1}{\beta\hbar \omega}}} , \quad  
\lim_{\beta\to 0} A_{\mathcal{C}}= \ln \frac{\omega_R^2 +\omega^2}{2\omega_R\, \omega}.
\end{equation}
The last limit shows that for very high temperature the amplitude of oscillations saturates at an upper bound (the black dot on Fig.\,\ref{Ac(b)}).

\subsubsection{Frequency analysis}
At fixed temperature we can consider complexity as a function of the 
cyclotron frequency (the magnetic field). For example, at high frequency ($\beta\hbar\omega\gg 1$ and $\omega/\omega_R=\delta \gg 1$) the asymptotic behavior of complexity \eqref{Complexity} follows from \eqref{red}:
\begin{equation}\label{Cwinfty}
\mathcal C(t)\approx \sqrt{2} \ln\delta + \frac{\ln^2 6}{2\sqrt{2} \ln \delta}, \quad \lim_{\omega\to\infty} \mathcal C(t)= \infty .
\end{equation}
Since $\delta\gg 1$ we have a logarithmic divergence at $\omega\to\infty$.
The amplitude at high frequency is the asymptotic expansion of \eqref{ALowT} with respect to $\delta\gg 1$:
\begin{equation}\label{ACwinfty}
A_{\mathcal C} \approx \sqrt{2}\, e^{-\beta \hbar\delta\omega_R } \bigg( 1 -\frac{1}{\ln\delta} \bigg),  \quad   \lim_{\omega\to \infty} A_{\mathcal{C}}= 0.
\end{equation}
Obviously the the amplitude is exponentially suppressed by $\delta$, hence it vanishes at $\omega\to\infty$. 

At low frequency ($\beta\hbar\omega\ll 1$ and $\omega/\omega_R=\delta\ll 1$) we use \eqref{HighTt}: 
\begin{equation}\label{Cw0}
{\cal C}(t) \approx   \ln\frac{1}{\beta\hbar\delta\omega_R} +\ln \frac{2}{\delta}\sqrt{\sin^2\omega t +2\delta^2 (1+\cos^2\omega t)}, \quad  \lim_{\omega\to 0} \mathcal C(t)= \infty .
\end{equation}
In this case complexity has a logarithmic  diverges at $\omega\to 0$.
This is also true for the amplitude of oscillations:
\begin{equation}\label{Acw0}
\lim_{\omega\to 0} A_{\mathcal{C}}= \infty.
\end{equation}
This behavior of the complexity for a quantum particle stands in apparent contrast to the harmonic oscillator case analyzed in \cite{Avramov:2024kpt}. In our scenario, the external magnetic field cannot be turned off, whereas in the harmonic oscillator case, the presence of an additional harmonic frequency allows for this possibility. Moreover, it can be shown that the complexity of the quantum particle cannot be derived as a limiting case of the harmonic oscillator.

\begin{figure}[H]
\centering \hspace{-1.0cm}
\begin{subfigure}{0.4\textwidth}
\includegraphics[width=8.3cm,height=5.55cm]{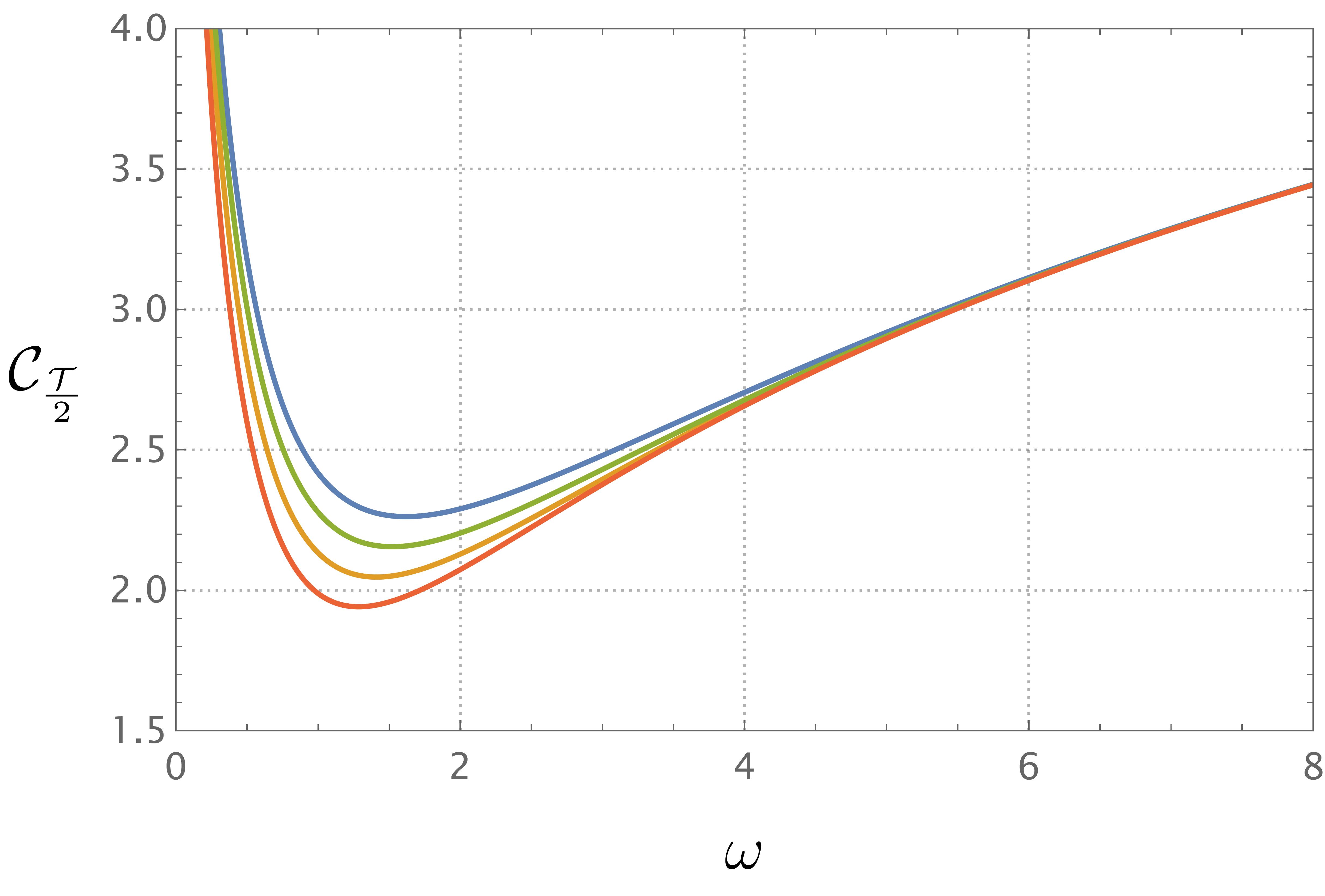}
\caption{Complexity $\mathcal C_{\frac{\mathcal{T}}{2}}(\omega)$.}\label{Cw}
\end{subfigure}
\hspace{1.5 cm}
\begin{subfigure}{0.4\textwidth}
\includegraphics[width=8.3cm,height=5.5cm]{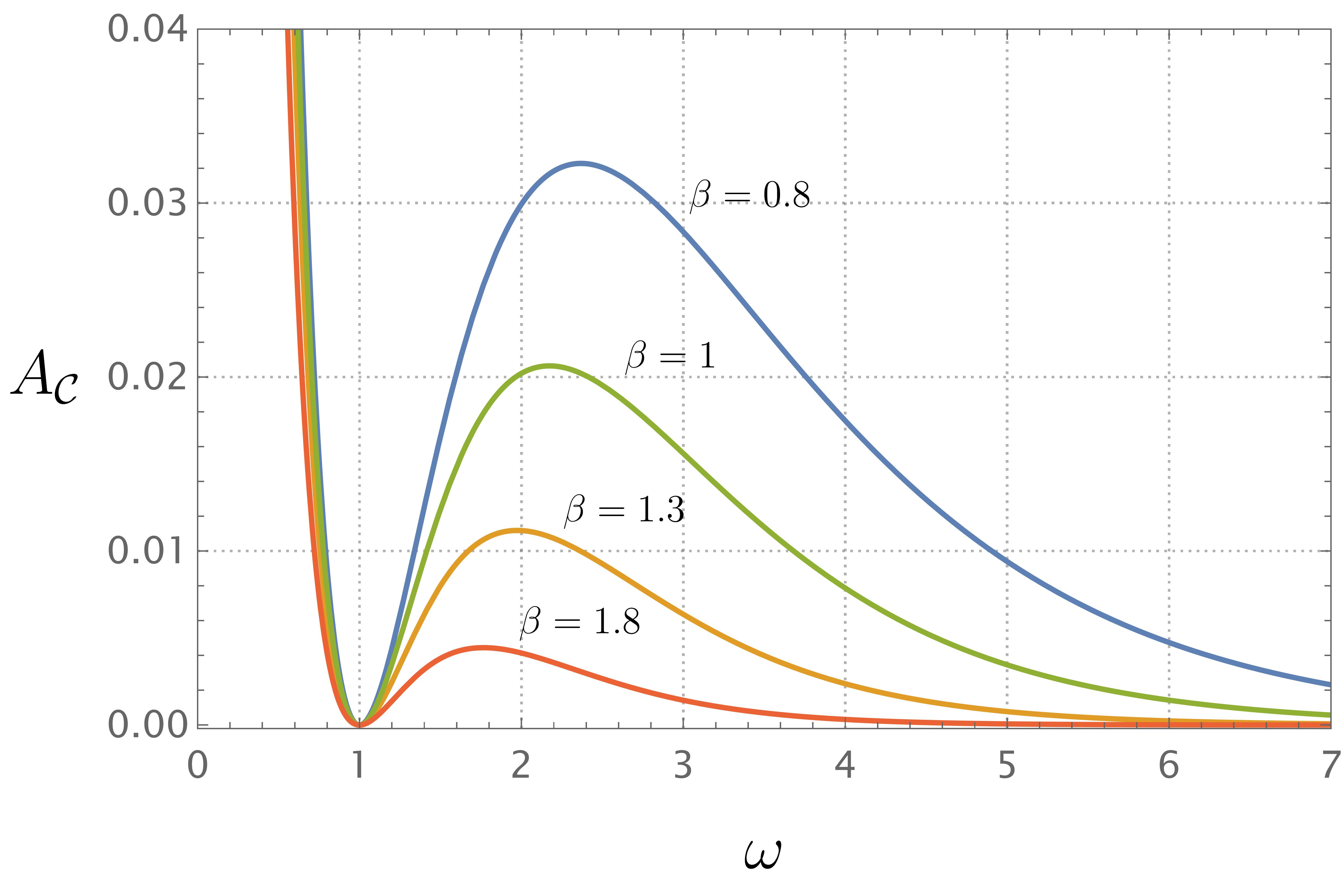}\vspace{-0.05cm}
\caption{Amplitude of oscillations $A_{\mathcal C}(\omega)$. }\label{Acw}
\end{subfigure}
 \hspace{0.7 cm}
\caption{ \textbf{(a)} Complexity $\mathcal C_{\frac{\mathcal{T}}{2}}(\omega)$ as a functions of the frequency at $t=\mathcal{T}/2$.
\textbf{(b)}  Amplitude of complexity oscillations $A_{\mathcal C}(\omega)$ as function of the frequency. We choose $\omega_R=1, \beta=1$ and $\hbar=1$.
}\label{figCwAcw}
\end{figure}

On Figure \ref{figCwAcw} we show the maximum value of complexity and the amplitude of oscillations as functions of the frequency at different values of the temperature. One notes the logarithmic divergence of the complexity for high and low frequency and the exponential suppression of the amplitude of oscillations for high frequency.  
\section{Rate of complexity and Lloyd's bound} \label{sec6}

\subsection{Internal energy}
The internal energy of the TFD state can be derived  from the partition function \eqref{PartF}: 
\begin{equation} \label{lloyd_bound}
U=-\,\frac{\partial \ln Z}{\partial \beta} = \frac{\hbar\omega}{2} \coth \frac{\beta\hbar\omega}{2},
\end{equation}
which coincides with the internal energy of a harmonic oscillator.
At zero temperature limit the internal energy reduces to the ground state energy \eqref{E_n}:
\begin{equation}\label{E00}
\lim_{\beta\to \infty} U=\frac{\hbar\omega}{2} =E_0.
\end{equation}
At high temperature (\(\beta \hbar \omega \ll 1\)) quantum effects become negligible and the internal energy asymptotically scales linearly with the temperature $U\approx 1/\beta=kT$. Given the internal energy, we can now calculate the system's rate of complexity. According to \cite{lloyd}, the rate of complexity must obey the energy-related limit known as the Lloyd bound. In the following, we demonstrate that the quantum charged particle complies with this bound.

\subsection{Rate of complexity}

The rate of complexity is the time derivative of \eqref{Complexity}:
\begin{equation}
\dot{\mathcal C}(t) =\frac{1}{4\mathcal C(t)} \sum_{s=1}^4 \frac{\dot e_s}{e_s} \ln e_s. 
\end{equation}
In the zero temperature limit one has $\lim\limits_{\beta\to \infty} \mathcal {\dot C}(t)= 0$, which corresponds to the red line on Fig.\,\ref{Rate}. At the high temperature limit the rate of complexity becomes
\begin{equation}\label{RateCb0}
\lim_{\beta\to 0}  \dot{\mathcal C}(t) =\frac{1}{2}  \frac{\omega (\omega_R^2 -\omega^2)^2 \sin 2\omega t}{(\omega_R^2+\omega^2)^2 -(\omega_R^2-\omega^2)^2 \cos^2\omega t},
\end{equation}
which is an oscillating function with a finite amplitude (blue curve on Fig.\,\ref{Rate}) and the same period of oscillations $\mathcal T=\pi/\omega$.

\subsection{Lloyd's bound}
\begin{figure}[H]
\centering \hspace{-1.0cm}
\includegraphics[width=8.3cm,height=5.55cm]{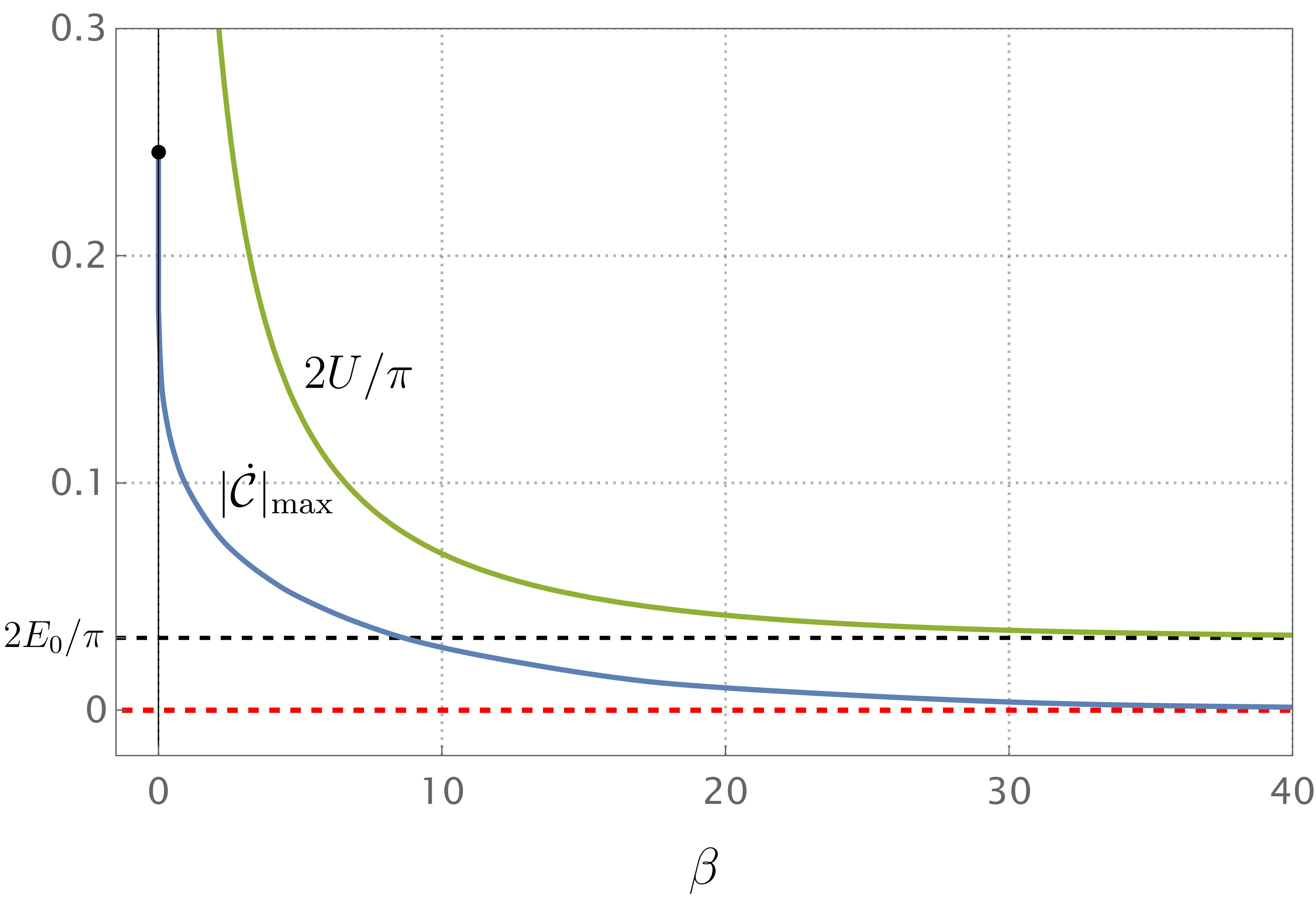}
\hspace{0.7 cm}
\caption{Maximum rate of complexity $|\dot{\mathcal C}|_{\rm max}$ (blue) and $2U/\pi$ (green curve) as functions of $\beta$. We choose: $\omega_R=1,\, \omega=0.1$ and $\hbar=1$. }\label{LloydBound}
\end{figure}

We aim to compare the rate of complexity with the system's internal energy, known as the Lloyd bound. It is defined within quantum information theory as \cite{lloyd}:
\begin{equation}\label{Lloyd}
|\dot{\mathcal C}|_{\rm{max}} \leq \frac{2U}{\pi\hbar}.
\end{equation}
Here \( |\mathcal{\dot{C}}|_{\rm{max}} \) is the amplitude of $\mathcal {\dot C}(t)$. We show that this bound \eqref{Lloyd} is satisfied for all temperatures as depicted on Fig.\,\ref{LloydBound}. In this case, \( |\mathcal{\dot{C}}|_{\rm{max}} \) (the blue curve) always remains below \(2U/\pi\) (the green curve). At low temperatures, the green curve approaches its minimum, which is proportional to the ground state energy (black dashed line), while the blue curve tends to zero (red dashed line). At high temperature, the green curve diverges, while the blue curve approaches the black dot, representing an upper limit of the rate of complexity. The later limit is the  amplitude of \eqref{RateCb0}.

\section{Conclusion}\label{sec7}

In this paper, we investigated the Nielsen complexity of a thermofield double state for a quantum charged particle subjected to an external magnetic field. We utilized the covariance matrix method, as described in \cite{chapman2019complexity}, which is particularly well-suited for examining the evolution of Gaussian states. Nielsen complexity, in general, quantifies the length of the optimal path connecting two quantum states, thereby providing a measure of the difficulty in transitioning between them. Based on these principles, we derived explicit results for both the complexity and its rate, and performed a comprehensive analysis across various temperatures and magnetic field strengths.

We investigated how complexity and the amplitude of its oscillations vary with temperature at a fixed cyclotron frequency, considering both low and high-temperature regimes. At low temperatures, the complexity approaches a minimum positive value \eqref{minValOfC}, and, as expected, the amplitude of oscillations is suppressed, tending toward zero \eqref{ALowT}. Conversely, at high temperatures, the complexity becomes unbounded \eqref{HighTt}, while the oscillation amplitude reaches a finite global maximum \eqref{AHighT}. Overall, increasing the temperature results in the growth of both complexity and its oscillations, as illustrated in Fig. \ref{figCb}.

A subsequent analysis of the frequency dependence of complexity and its amplitude at a fixed temperature reveals the following outcomes. At high frequencies, the complexity becomes unbounded \eqref{Cwinfty}, while the oscillation amplitude vanishes \eqref{ACwinfty}. Conversely, at low frequencies, both the complexity and its oscillation amplitude increase without bound \eqref{Cw0} and \eqref{Acw0}. This behavior is illustrated in Fig. \ref{figCwAcw}. Additionally, a comparison with the harmonic oscillator case \cite{Avramov:2024kpt} shows that the complexity of the quantum particle cannot be obtained as a limiting case of the harmonic oscillator setup. 

Finally, we investigated the Lloyd bound of the system, which links its internal energy to the maximum rate of complexity. Our findings demonstrate that, regardless of temperature, the rate of complexity consistently obeys this bound, as shown in Fig. \ref{LloydBound}. 

Although this study mainly focused on the impact of an external magnetic field  and the temperature on the complexity of a quantum particle, a similar approach could be extended to more general systems.

\section*{Acknowledgments}
The author would like to express his gratitude to R. C. Rashkov, T. Vetsov, V. Avramov, and H. Dimov for their invaluable comments
and discussions. M. Radomirov was fully financed by the European Union-Next Generation EU, through the National Recovery and Resilience Plan of the Republic
of Bulgaria, grant number BG-RRP-2.004-0008-C01.

\section*{Data availability}
This manuscript has no associated data or the data will not be deposited. All data generated or analyzed during this study are included in this published article.

\bibliographystyle{utphys}
\bibliography{ref}

\providecommand{\href}[2]{#2}\begingroup\raggedright\begin{thebibliography}{10}

\bibitem{Ryu:2006bv}
S.~Ryu and T.~Takayanagi, ``{Holographic derivation of entanglement entropy
  from AdS/CFT},'' \href{http://dx.doi.org/10.1103/PhysRevLett.96.181602}{{\em
  Phys. Rev. Lett.} {\bfseries 96} (2006) 181602},
  \href{http://arxiv.org/abs/hep-th/0603001}{{\ttfamily arXiv:hep-th/0603001}}.

\bibitem{Ryu:2006ef}
S.~Ryu and T.~Takayanagi, ``{Aspects of Holographic Entanglement Entropy},''
  \href{http://dx.doi.org/10.1088/1126-6708/2006/08/045}{{\em JHEP} {\bfseries
  08} (2006) 045}, \href{http://arxiv.org/abs/hep-th/0605073}{{\ttfamily
  arXiv:hep-th/0605073}}.

\bibitem{Susskind:2014moa}
L.~Susskind, ``{Entanglement is not enough},''
  \href{http://dx.doi.org/10.1002/prop.201500095}{{\em Fortsch. Phys.}
  {\bfseries 64} (2016) 49--71},
  \href{http://arxiv.org/abs/1411.0690}{{\ttfamily arXiv:1411.0690 [hep-th]}}.

\bibitem{Susskind:2014rva}
L.~Susskind, ``{Computational Complexity and Black Hole Horizons},''
  \href{http://dx.doi.org/10.1002/prop.201500092}{{\em Fortsch. Phys.}
  {\bfseries 64} (2016) 24--43},
  \href{http://arxiv.org/abs/1403.5695}{{\ttfamily arXiv:1403.5695 [hep-th]}}.
  [Addendum: Fortsch.Phys. 64, 44--48 (2016)].

\bibitem{Maldacena:2001kr}
J.~M. Maldacena, ``{Eternal black holes in anti-de Sitter},''
  \href{http://dx.doi.org/10.1088/1126-6708/2003/04/021}{{\em JHEP} {\bfseries
  04} (2003) 021}, \href{http://arxiv.org/abs/hep-th/0106112}{{\ttfamily
  arXiv:hep-th/0106112}}.

\bibitem{Hartman:2013qma}
T.~Hartman and J.~Maldacena, ``{Time Evolution of Entanglement Entropy from
  Black Hole Interiors},''
  \href{http://dx.doi.org/10.1007/JHEP05(2013)014}{{\em JHEP} {\bfseries 05}
  (2013) 014}, \href{http://arxiv.org/abs/1303.1080}{{\ttfamily arXiv:1303.1080
  [hep-th]}}.

\bibitem{Brown:2015bva}
A.~R. Brown, D.~A. Roberts, L.~Susskind, B.~Swingle, and Y.~Zhao,
  ``{Holographic Complexity Equals Bulk Action?},''
  \href{http://dx.doi.org/10.1103/PhysRevLett.116.191301}{{\em Phys. Rev.
  Lett.} {\bfseries 116} no.~19, (2016) 191301},
  \href{http://arxiv.org/abs/1509.07876}{{\ttfamily arXiv:1509.07876
  [hep-th]}}.

\bibitem{Couch:2016exn}
J.~Couch, W.~Fischler, and P.~H. Nguyen, ``{Noether charge, black hole volume,
  and complexity},'' \href{http://dx.doi.org/10.1007/JHEP03(2017)119}{{\em
  JHEP} {\bfseries 03} (2017) 119},
  \href{http://arxiv.org/abs/1610.02038}{{\ttfamily arXiv:1610.02038
  [hep-th]}}.

\bibitem{Belin:2021bga}
A.~Belin, R.~C. Myers, S.-M. Ruan, G.~S\'arosi, and A.~J. Speranza, ``{Does
  Complexity Equal Anything?},''
  \href{http://dx.doi.org/10.1103/PhysRevLett.128.081602}{{\em Phys. Rev.
  Lett.} {\bfseries 128} no.~8, (2022) 081602},
  \href{http://arxiv.org/abs/2111.02429}{{\ttfamily arXiv:2111.02429
  [hep-th]}}.

\bibitem{Belin:2022xmt}
A.~Belin, R.~C. Myers, S.-M. Ruan, G.~S\'arosi, and A.~J. Speranza,
  ``{Complexity equals anything II},''
  \href{http://dx.doi.org/10.1007/JHEP01(2023)154}{{\em JHEP} {\bfseries 01}
  (2023) 154}, \href{http://arxiv.org/abs/2210.09647}{{\ttfamily
  arXiv:2210.09647 [hep-th]}}.

\bibitem{Jorstad:2023kmq}
E.~J\o{}rstad, R.~C. Myers, and S.-M. Ruan, ``{Complexity=anything: singularity
  probes},'' \href{http://dx.doi.org/10.1007/JHEP07(2023)223}{{\em JHEP}
  {\bfseries 07} (2023) 223}, \href{http://arxiv.org/abs/2304.05453}{{\ttfamily
  arXiv:2304.05453 [hep-th]}}.

\bibitem{Myers:2024vve}
R.~C. Myers and S.-M. Ruan, ``{Complexity Equals (Almost) Anything},''
  \href{http://arxiv.org/abs/2403.17475}{{\ttfamily arXiv:2403.17475
  [hep-th]}}.

\bibitem{arora2009computational}
S.~Arora and B.~Barak, {\em Computational Complexity: A Modern Approach}.
\newblock Cambridge University Press, 2009.
\newblock \url{https://books.google.de/books?id=8Wjqvsoo48MC}.

\bibitem{moore2011nature}
C.~Moore and S.~Mertens, {\em The Nature of Computation}.
\newblock OUP Oxford, 2011.
\newblock \url{https://books.google.de/books?id=jnGKbpMV8xoC}.

\bibitem{Nielsen:2005mkt}
M.~A. Nielsen, ``{A geometric approach to quantum circuit lower bounds},''
  \href{http://dx.doi.org/10.26421/QIC6.3-2}{{\em Quant. Inf. Comput.}
  {\bfseries 6} no.~3, (2006) 213--262},
  \href{http://arxiv.org/abs/quant-ph/0502070}{{\ttfamily
  arXiv:quant-ph/0502070}}.

\bibitem{nielsen2010quantum}
M.~Nielsen and I.~Chuang, {\em Quantum Computation and Quantum Information:
  10th Anniversary Edition}.
\newblock Cambridge University Press, 2010.
\newblock \url{https://books.google.de/books?id=-s4DEy7o-a0C}.

\bibitem{Dowling:2006tnk}
M.~R. Dowling and M.~A. Nielsen, ``{The geometry of quantum computation},''
  \href{http://dx.doi.org/10.26421/QIC8.10-1}{{\em Quant. Inf. Comput.}
  {\bfseries 8} no.~10, (2008) 0861--0899},
  \href{http://arxiv.org/abs/quant-ph/0701004}{{\ttfamily
  arXiv:quant-ph/0701004}}.

\bibitem{Nielsen:2006cea}
M.~A. Nielsen, M.~R. Dowling, M.~Gu, and A.~C. Doherty, ``{Quantum Computation
  as Geometry},'' \href{http://dx.doi.org/10.1126/science.1121541}{{\em
  Science} {\bfseries 311} no.~5764, (2006) 1133--1135},
  \href{http://arxiv.org/abs/quant-ph/0603161}{{\ttfamily
  arXiv:quant-ph/0603161}}.

\bibitem{cormen2001introduction}
T.~Cormen, C.~Leiserson, R.~Rivest, and C.~Stein, {\em Introduction To
  Algorithms}.
\newblock Mit Electrical Engineering and Computer Science. MIT Press, 2001.
\newblock \url{https://books.google.de/books?id=NLngYyWFl_YC}.

\bibitem{Chapman:2018hou}
S.~Chapman, J.~Eisert, L.~Hackl, M.~P. Heller, R.~Jefferson, H.~Marrochio, and
  R.~C. Myers, ``{Complexity and entanglement for thermofield double states},''
  \href{http://dx.doi.org/10.21468/SciPostPhys.6.3.034}{{\em SciPost Phys.}
  {\bfseries 6} no.~3, (2019) 034},
  \href{http://arxiv.org/abs/1810.05151}{{\ttfamily arXiv:1810.05151
  [hep-th]}}.

\bibitem{Ghasemi:2021jiy}
M.~Ghasemi, A.~Naseh, and R.~Pirmoradian, ``{Odd entanglement entropy and
  logarithmic negativity for thermofield double states},''
  \href{http://dx.doi.org/10.1007/JHEP10(2021)128}{{\em JHEP} {\bfseries 10}
  (2021) 128}, \href{http://arxiv.org/abs/2106.15451}{{\ttfamily
  arXiv:2106.15451 [hep-th]}}.

\bibitem{Doroudiani:2019llj}
M.~Doroudiani, A.~Naseh, and R.~Pirmoradian, ``{Complexity for Charged
  Thermofield Double States},''
  \href{http://dx.doi.org/10.1007/JHEP01(2020)120}{{\em JHEP} {\bfseries 01}
  (2020) 120}, \href{http://arxiv.org/abs/1910.08806}{{\ttfamily
  arXiv:1910.08806 [hep-th]}}.

\bibitem{Khorasani:2023usq}
F.~Khorasani, R.~Pirmoradian, and M.~R. Tanhayi, ``{Position dependence of
  Nielsen complexity for the thermofield double state},''
  \href{http://dx.doi.org/10.1016/j.physletb.2024.138585}{{\em Phys. Lett. B}
  {\bfseries 851} (2024) 138585},
  \href{http://arxiv.org/abs/2308.15836}{{\ttfamily arXiv:2308.15836
  [quant-ph]}}.

\bibitem{Chapman:2017rqy}
S.~Chapman, M.~P. Heller, H.~Marrochio, and F.~Pastawski, ``{Toward a
  Definition of Complexity for Quantum Field Theory States},''
  \href{http://dx.doi.org/10.1103/PhysRevLett.120.121602}{{\em Phys. Rev.
  Lett.} {\bfseries 120} no.~12, (2018) 121602},
  \href{http://arxiv.org/abs/1707.08582}{{\ttfamily arXiv:1707.08582
  [hep-th]}}.

\bibitem{Bhattacharyya:2018bbv}
A.~Bhattacharyya, A.~Shekar, and A.~Sinha, ``{Circuit complexity in interacting
  QFTs and RG flows},'' \href{http://dx.doi.org/10.1007/JHEP10(2018)140}{{\em
  JHEP} {\bfseries 10} (2018) 140},
  \href{http://arxiv.org/abs/1808.03105}{{\ttfamily arXiv:1808.03105
  [hep-th]}}.

\bibitem{Ali:2018fcz}
T.~Ali, A.~Bhattacharyya, S.~Shajidul~Haque, E.~H. Kim, and N.~Moynihan,
  ``{Time Evolution of Complexity: A Critique of Three Methods},''
  \href{http://dx.doi.org/10.1007/JHEP04(2019)087}{{\em JHEP} {\bfseries 04}
  (2019) 087}, \href{http://arxiv.org/abs/1810.02734}{{\ttfamily
  arXiv:1810.02734 [hep-th]}}.

\bibitem{Jefferson:2017sdb}
R.~Jefferson and R.~C. Myers, ``{Circuit complexity in quantum field theory},''
  \href{http://dx.doi.org/10.1007/JHEP10(2017)107}{{\em JHEP} {\bfseries 10}
  (2017) 107}, \href{http://arxiv.org/abs/1707.08570}{{\ttfamily
  arXiv:1707.08570 [hep-th]}}.

\bibitem{Bagchi:2001dx}
B.~Bagchi, {\em Supersymmetry In Quantum and Classical Mechanics}.
\newblock Chapman \& Hall/CRC monographs and surveys in pure and applied
  mathematics. Taylor \& Francis Limited (Sales), 2019.

\bibitem{Jafarov:2013cza}
E.~I. Jafarov and J.~Van~der Jeugt, ``{The oscillator model for the Lie
  superalgebra sh(2|2) and Charlier polynomials},''
  \href{http://dx.doi.org/10.1063/1.4824742}{{\em J. Math. Phys.} {\bfseries
  54} (2013) 103506}, \href{http://arxiv.org/abs/1304.3295}{{\ttfamily
  arXiv:1304.3295 [math-ph]}}.

\bibitem{Fatyga:1990wx}
B.~W. Fatyga, V.~A. Kostelecky, M.~M. Nieto, and D.~R. Truax, ``{Supercoherent
  states},'' \href{http://dx.doi.org/10.1103/PhysRevD.43.1403}{{\em Phys. Rev.
  D} {\bfseries 43} (1991) 1403--1412}.

\bibitem{Mandal:2012wp}
B.~P. Mandal and S.~K. Rai, ``{Noncommutative Dirac oscillator in an external
  magnetic field},''
  \href{http://dx.doi.org/10.1016/j.physleta.2012.07.001}{{\em Phys. Lett. A}
  {\bfseries 376} (2012) 2467--2470},
  \href{http://arxiv.org/abs/1203.2714}{{\ttfamily arXiv:1203.2714 [hep-th]}}.

\bibitem{Jing:2009lfc}
J.~Jing and J.-F. Chen, ``{Non-commutative harmonic oscillator in magnetic
  field and continuous limit},''
  \href{http://dx.doi.org/10.1140/epjc/s10052-009-0950-1}{{\em Eur. Phys. J. C}
  {\bfseries 60} (2009) 669--674}.

\bibitem{BenGeloun:2009hkc}
J.~Ben~Geloun, S.~Gangopadhyay, and F.~G. Scholtz, ``{Harmonic oscillator in a
  background magnetic field in noncommutative quantum phase-space},''
  \href{http://dx.doi.org/10.1209/0295-5075/86/51001}{{\em EPL} {\bfseries 86}
  no.~5, (2009) 51001}, \href{http://arxiv.org/abs/0901.3412}{{\ttfamily
  arXiv:0901.3412 [hep-th]}}.

\bibitem{Heddar:2021sly}
M.~Heddar, M.~Falek, M.~Moumni, and B.~C. L\"utf\"uo\u{g}lu, ``{Pauli
  oscillator in noncommutative space},''
  \href{http://dx.doi.org/10.1142/S0217732321502801}{{\em Mod. Phys. Lett. A}
  {\bfseries 36} no.~40, (2021) 2150280},
  \href{http://arxiv.org/abs/2110.00723}{{\ttfamily arXiv:2110.00723
  [hep-th]}}.

\bibitem{Boumali:2020fqd}
A.~Boumali, F.~Serdouk, and S.~Dilmi, ``{Superstatistical properties of the
  one-dimensional Dirac oscillator},''
  \href{http://dx.doi.org/10.1016/j.physa.2020.124207}{{\em Physica A}
  {\bfseries 553} (2020) 124207}.

\bibitem{Falek:2017amp}
M.~Falek, M.~Merad, and T.~Birkandan,
  ``{Duffin\textendash{}Kemmer\textendash{}Petiau oscillator with Snyder-de
  Sitter algebra},'' \href{http://dx.doi.org/10.1063/1.4975137}{{\em J. Math.
  Phys.} {\bfseries 58} no.~2, (2017) 023501}.

\bibitem{Nagiyev:2023fwk}
S.~M. Nagiyev and R.~M. Mir-Kasimov, ``{Relativistic linear oscillator under
  the action of a constant external force. Transition amplitudes and the
  Green\textquoteright{}s function},''
  \href{http://dx.doi.org/10.1134/S004057792301004X}{{\em Theor. Math. Phys.}
  {\bfseries 214} no.~1, (2023) 72--88}.

\bibitem{Martinez-y-Romero:1995grr}
R.~P. Martinez-y Romero, H.~N. Nunez-Yepez, and A.~L. Salas-Brito,
  ``{Relativistic quantum mechanics of a Dirac oscillator},''
  \href{http://dx.doi.org/10.1088/0143-0807/16/3/008}{{\em Eur. J. Phys.}
  {\bfseries 16} (1995) 135--141},
  \href{http://arxiv.org/abs/quant-ph/9908069}{{\ttfamily
  arXiv:quant-ph/9908069}}.

\bibitem{Mannheim:2004qz}
P.~D. Mannheim and A.~Davidson, ``{Dirac quantization of the Pais-Uhlenbeck
  fourth order oscillator},''
  \href{http://dx.doi.org/10.1103/PhysRevA.71.042110}{{\em Phys. Rev. A}
  {\bfseries 71} (2005) 042110},
  \href{http://arxiv.org/abs/hep-th/0408104}{{\ttfamily arXiv:hep-th/0408104}}.

\bibitem{Masterov:2015ija}
I.~Masterov, ``{An alternative Hamiltonian formulation for the
  Pais\textendash{}Uhlenbeck oscillator},''
  \href{http://dx.doi.org/10.1016/j.nuclphysb.2015.11.011}{{\em Nucl. Phys. B}
  {\bfseries 902} (2016) 95--114},
  \href{http://arxiv.org/abs/1505.02583}{{\ttfamily arXiv:1505.02583
  [hep-th]}}.

\bibitem{Dimov:2016vvl}
H.~Dimov, S.~Mladenov, R.~C. Rashkov, and T.~Vetsov, ``{Entanglement of
  higher-derivative oscillators in holographic systems},''
  \href{http://dx.doi.org/10.1016/j.nuclphysb.2017.03.005}{{\em Nucl. Phys. B}
  {\bfseries 918} (2017) 317--336},
  \href{http://arxiv.org/abs/1607.07807}{{\ttfamily arXiv:1607.07807
  [hep-th]}}.

\bibitem{Pramanik:2012bh}
S.~Pramanik and S.~Ghosh, ``{Taming the Ghost in Pais-Uhlenbeck Oscillator},''
  \href{http://dx.doi.org/10.1142/S0217732313500387}{{\em Mod. Phys. Lett. A}
  {\bfseries 28} (2013) 1350038},
  \href{http://arxiv.org/abs/1205.3333}{{\ttfamily arXiv:1205.3333 [math-ph]}}.

\bibitem{Bouguerne:2023wyl}
H.~Bouguerne, B.~Hamil, B.~C. L\"utf\"uo\u{g}lu, and M.~Merad, ``{Dunkl-Pauli
  Equation in the Presence of a Magnetic Field},''
  \href{http://arxiv.org/abs/2309.14081}{{\ttfamily arXiv:2309.14081
  [quant-ph]}}.

\bibitem{Hamil:2022uwy}
B.~Hamil and B.~C. L\"utf\"uo\u{g}lu, ``{Thermal properties of relativistic
  Dunkl oscillators},''
  \href{http://dx.doi.org/10.1140/epjp/s13360-022-03055-1}{{\em Eur. Phys. J.
  Plus} {\bfseries 137} no.~7, (2022) 812},
  \href{http://arxiv.org/abs/2202.02871}{{\ttfamily arXiv:2202.02871
  [quant-ph]}}.

\bibitem{Ballesteros:2022bqx}
A.~Ballesteros, A.~Najafizade, H.~Panahi, H.~Hassanabadi, and S.-H. Dong,
  ``{The Dunkl oscillator on a space of nonconstant curvature: An exactly
  solvable quantum model with reflections},''
  \href{http://dx.doi.org/10.1016/j.aop.2023.169543}{{\em Annals Phys.}
  {\bfseries 460} (2024) 169543},
  \href{http://arxiv.org/abs/2212.13575}{{\ttfamily arXiv:2212.13575
  [quant-ph]}}.

\bibitem{chapman2019complexity}
S.~Chapman, J.~Eisert, L.~Hackl, M.~P. Heller, R.~Jefferson, H.~Marrochio, and
  R.~Myers, ``Complexity and entanglement for thermofield double states,'' {\em
  SciPost physics} {\bfseries 6} no.~3, (2019) 034.

\bibitem{Avramov:2024kpt}
V.~Avramov, M.~Radomirov, R.~C. Rashkov, and T.~Vetsov, ``{Complexity of
  Quantum Harmonic Oscillator in External Magnetic Field},''
  \href{http://arxiv.org/abs/2407.18631}{{\ttfamily arXiv:2407.18631
  [quant-ph]}}.

\bibitem{lloyd}
S.~Lloyd, ``Ultimate physical limits to computation,'' {\em Nature} {\bfseries
  406} no.~6799, (2000) 1047--1054.

\end{thebibliography}\endgroup

\end{document}